# Automated Segmentation of Large Image Datasets using Artificial Intelligence for Microstructure Characterisation, Damage Analysis and High-Throughput Modelling Input

Authors

Setareh Medghalchi[1,*], Joscha Kortmann[1], Sang-Hyeok Lee[1], Ehsan Karimi[1], Ulrich Kerzel[2], Sandra Korte-Kerzel[1]

Affiliation(s)

1. Institute for Physical Metallurgy and Materials Physics, RWTH Aachen University, Aachen, Germany
2. Data Science and Artificial Intelligence in Materials and Geoscience, Fakultät für Georessourcen und Materialtechnik, RWTH Aachen University, Aachen, Germany

*Corresponding author

## Graphical Abstract:

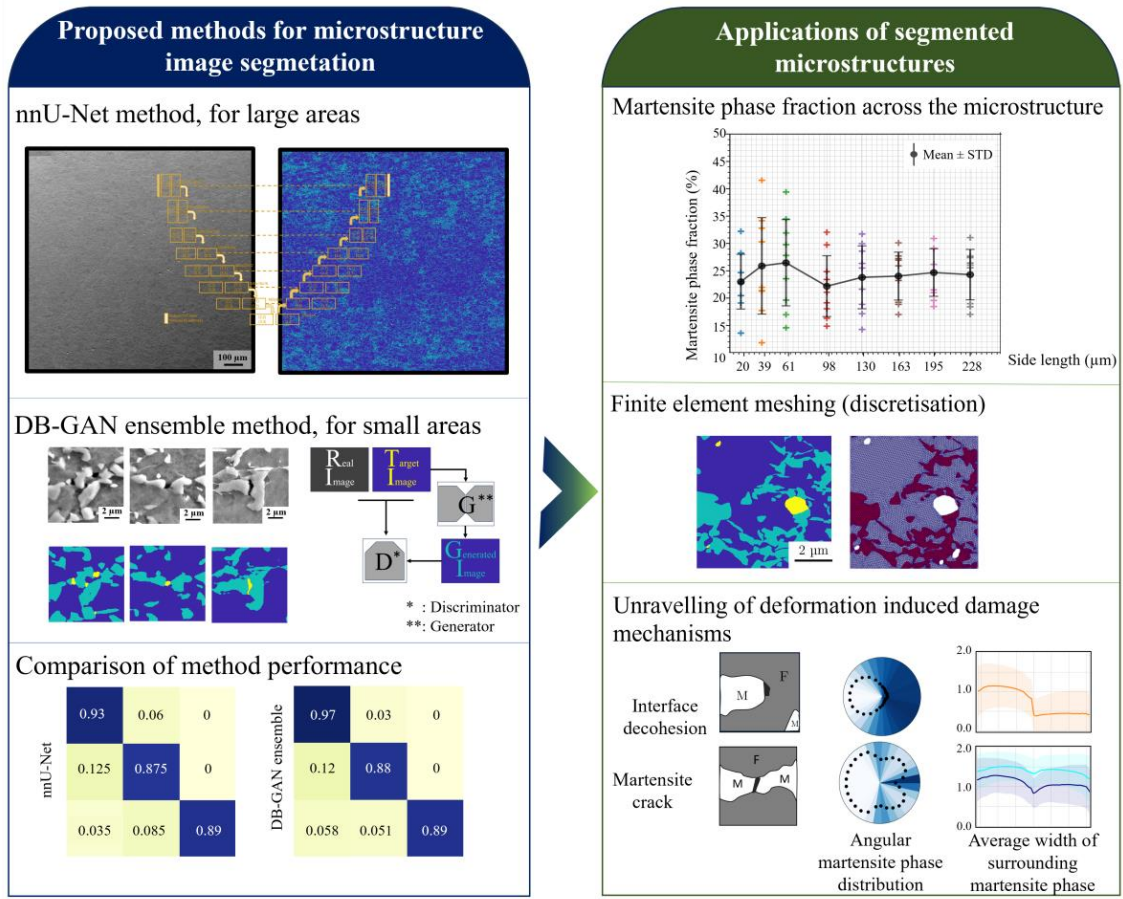

# Abstract

Many properties of commonly used materials are driven by their microstructure, which can be influenced by the composition and manufacturing processes. To optimise future materials, understanding the microstructure is critically important. Here, we present two novel approaches based on artificial intelligence that allow the segmentation of the phases of a microstructure for which simple numerical approaches, such as thresholding, are not applicable: One is based on the nnU-Net neural network, and the other on generative adversarial networks (GAN).

Using large panoramic scanning electron microscopy images of dual-phase steels as a case study, we demonstrate how both methods effectively segment intricate microstructural details, including martensite, ferrite, and damage sites, for subsequent analysis.

Either method shows substantial generalizability across a range of image sizes and conditions, including heat-treated microstructures with different phase configurations. The nnU-Net excels in mapping large image areas. Conversely, the GAN-based method performs reliably on smaller images, providing greater step-by-step control and flexibility over the segmentation process.

This study highlights the benefits of segmented microstructural data for various purposes, such as calculating phase fractions, modelling material behaviour through finite element simulation, and conducting geometrical analyses of damage sites and the local properties of their surrounding microstructure.

# 1 Introduction

Almost all materials used in today's applications are produced and tailored using specific production processes, meaning that their properties do not only depend on the chemical composition, but, in particular, the microstructure that is formed from the various phases, their size and morphology, crystal orientations and lattice defects in the material. Much of materials science research is, therefore, focused on understanding the properties that emerge from the microstructure and how to develop tailored processing chains to obtain materials with well-controlled microstructural variables yielding desirable properties. To be able to understand the relationship between the microstructure and the properties of the material quantitatively, one continuing challenge is to accurately describe the microstructure regarding the present phases and phase fractions as well as their geometries [1]. Such a representation not only enables comparisons of different materials or assessment of processing success, it also forms the basis of computational analyses or studies of microstructure evolution during straining, heat treatment [2], welding [3], or other exposure to different environments [4].

The domain of microstructure modelling and design typically uses a limited set of variables or descriptors to represent the microstructure on three levels: i) composition (comprising phases and their volume fractions), ii) dispersion (discriminating between distinct microstructural components), and iii) particle phase geometry (can include phase attributes like size, area, and roundness) [1]. However, it is important to keep a balance as excessive simplification by limiting the number of descriptors or uncertainty in their quantification may lead to loss of physical significance [4]. In the past century, materials science has made substantial progress in acquiring and analysing microstructural images to provide as detailed as possible a view of each material's microstructure [5, 6]. Microstructural image segmentation emerges as a pivotal technique, enabling the dissection of intricate visual data into meaningful components [7]. Microstructural image segmentation commonly involves partitioning of a microstructure image into its constituent parts, such as grains [8-13], phases, and defects [14-16], assigning physically meaningful and spatially structured data to its constituents [17, 18]. This also enables a direct transfer of experimental data into modelling of microstructural processes or properties, forming a bridge between theory and experiment.

Conventionally, microstructure segmentation is performed manually by human experts and supported by a range of numerical approaches such as thresholding [19-22], region growing [23-25], edge detection [26, 27], and clustering [28]. In addition, advancing analytical methods assist to automate this process, for example by using correlated micrographs with different electron detectors, giving topographical or chemical contrast, or employing energy dispersive spectroscopy (EDS) or electron backscatter diffraction (EBSD) inside the scanning electron microscopy (SEM) to more explicitly distinguished phases and grains. However, this type of analysis, while providing a wealth of information and high spatial resolution, is again time- and resource-consuming and rarely scaled to areas representative of a finished or semi-finished product. Lately, machine learning and increasingly automated imaging equipment are emerging to bridge this gap in scale [29, 30]. These methods enable analysis of large datasets, such as from synchrotron tomography [31-33], X-ray tomography [34], micro-computer tomography (CT) [32] and optical or electron micrographs [33, 35, 36], and have been adopted for different purposes including feature detection and classification [29, 30, 37, 38] and microstructure segmentation [32, 34-36, 39, 40]. Using images recorded at different scales, from light optical microscopy (LOM) to high resolution transmission electron microscopy (HR-TEM), objects or phases of interest have been segmented in various microstructures, for example, including ferrite, martensite, cementite particles and lath shaped bainite in different grades of steels [39, 41-45], or γ′ precipitates, dendrites [34] and their cores [46] in nickel based superalloys [47-49].

However, most of these methods are not suitable for analysing large areas. Typical challenges include handling a large dataset itself or dealing with variability in the underlying data as a result of changes in focus, contrast or brightness during imaging or fluctuations of the surface quality after metallographic preparation.

In this work, we present two novel approaches for microstructure segmentation based on artificial intelligence, one method is based on generative adversarial neural networks (GAN), and the other on nnU-Net. We employ both methods on panoramic scanning electron microscopy images revealing features of sub-micron scale across images that span hundreds of micrometres (μm) in either direction. As our sample material, we image a commercial dual phase steel (DP800). This type of steel is used extensively in a wide range of commercial applications, particularly in the automotive industry [50, 51], and improvements in this material have direct impact on many production processes and their applications [2, 7, 52, 53]. It also has a rich multi-phase microstructure that is formed during sheet processing and tailored to sustain large mechanical strains before failure. The final microstructure of the as-processed or subsequently strained material contains features ranging from small voids reaching the resolution limit of scanning electron microscopy to large martensite bands undulating at a shallow angle along the sheet material and giving rise to strong variations in the microstructure observed in sections taken parallel to the sheet itself [30, 38]. Furthermore, unlike many other materials containing multiple phases [54], the two major constituent phases of the dual phase steel, the brittle martensite and much more ductile ferrite phase [55], appear very similar in scanning electron microscopy and therefore provide an adequate challenge for segmentation where simple thresholding, for example, fails across large images.

In addition to a brief introduction to the two segmentation methods employed here and their application to the dual phase DP800 steel, we will present examples of how segmented microstructural data of large images can be used for subsequent analyses. Three aspects will be considered in these examples: (1) the apparently simple task of determining the phase fraction across a microstructure of large variability at different scales, (2) the use of segmentation to distinguish strain-induced voids or extract core geometric features after manual or neural network-based classification, and (3) the modelling of the microstructure's deformation once unambiguous phase assignments from segmented micrographs. A summary of the approaches and examples in this work is illustrated in Figure 1.

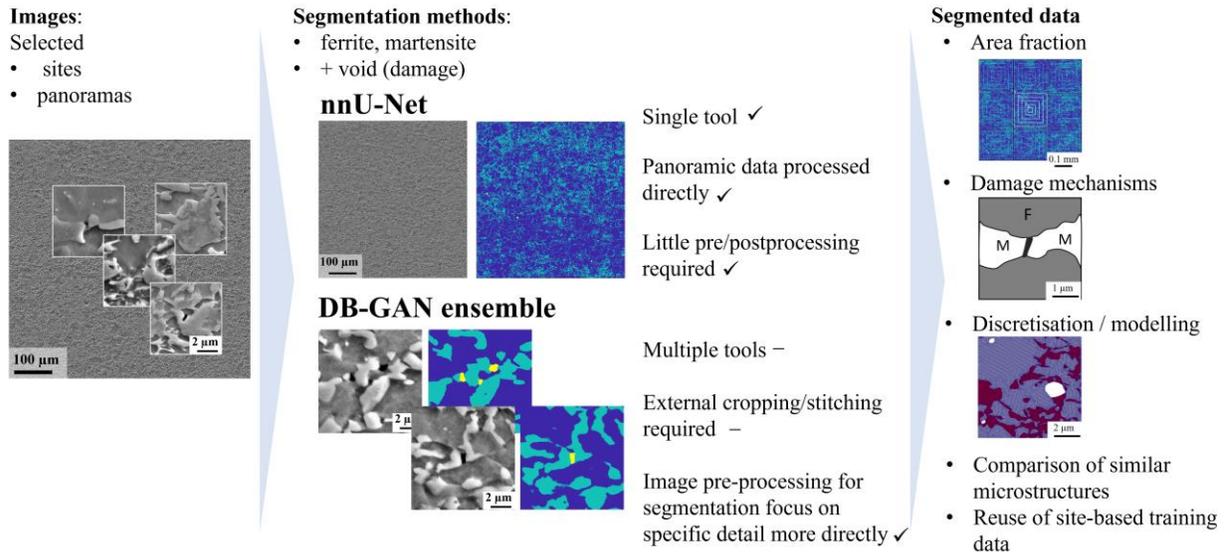

*Figure 1: Two approaches for automatic segmentation of the microstructural images with the corresponding benefits of segmented data.*

## 2 Introduction to the implemented segmentation methods and materials

### 2.1 Artificial Intelligence for Image Segmentation

In order to analyse a microstructure in detail, we need to be able to attribute each part of the micrograph to a specific feature of the image, in our case either a martensite or ferrite phase, or a void introduced as damage during mechanical straining. On an abstract level, we, therefore, want to use a machine learning algorithm to "translate" an input image to a suitable output image showing which phase or damage is where. In our case of the DP800 steel, the input image is the grayscale electron micrograph obtained from a scanning electron microscope, and the desired output image shows, pixel by pixel, whether a given pixel originates from the martensite phase, the ferrite phase, or a damage site. This process is called segmentation, and we distinguish between instance and semantic segmentation: In semantic segmentation, we attribute each pixel in an image to a pre-defined class, such as martensite, ferrite, or damage. Instance segmentation goes beyond semantic segmentation and aims to identify each object uniquely with its associated segmented pixels, such that each object can be referred to separately [56, 57]. In this work, we focus on semantic segmentation and will refer to semantic segmentation and segmentation synonymously throughout the text.

The training data for the machine learning based approaches used in this work were obtained by manually labelling scanning electron micrographs, using the Pixel label option of the Image Labeler application of MATLAB 2018a [59]. This tool provided the most precise labelling possibility at the time of preparing this data. Figure 2 shows an example of the raw (left) and labelled (right) image. Each image in the training data has a size of 512 × 512 pixels that corresponds to a resolution of 32 pixels/μm. In total, we have labelled 171 SEM images. The labelling of the ground truth data primarily serves the purpose of defining the phases. It involves the manual process of successive clicking around specified phases, namely martensite and damage sites, within the ferrite matrix. However, certain minute details, such as extremely thin martensite islands or remnants of OPS particles resulting from the metallographic preparation, that lack significant physical relevance to the material's properties, are excluded from this process to alleviate the effort. The orange arrows in Figure 2 indicate a few of these cases.

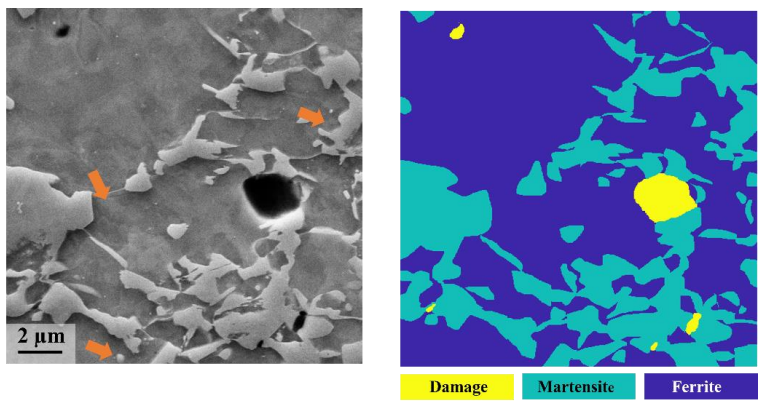

*Figure 2: Example of the manually labelled ground truth data with different colours for ferrite, martensite and damage sites. Orange arrows indicate some minute details which have not been manually marked.*

### 2.1.1 nnU-Net

The UNet architecture of convolutional neural networks [60] was primarily designed for image segmentation in biomedical images. It consists of a contraction part based on convolutional neural networks that learn the abstract features in the image, a "bridge", and an expansive part that then creates the output image. So called "skip connections" between the contraction and expansive parts convey information about specific features across the network topology. The name UNet originates from the U-like shape of the setup after the skip-connections between the contraction and expansive part are introduced [61]. This approach has been widely used for various computer vision tasks. Its efficient use of parameters allows the training on a limited number of training data or devices with few resources [58, 60, 62, 63].

The most recent variant is nnU-Net ("no new UNet") [62], which extends the original UNet approach to include automatic configuration, including the pre- and postprocessing of the data, the network architecture, and the network training, thereby alleviating many of the difficult and tedious tuning steps required in most neural network-based approaches. The postprocessing can include, for example, the removal of artifacts or smoothing of edges [64, 65]. To avoid the risk in our application of this method for the analysis of dual phase steel that damage sites or small martensite island may be removed, we decided not to apply any post-processing in this work.

The nnU-Net approach uses three classes of parameters: Fixed parameters, rule-based parameters, and empirical parameters. The fixed parameters have been tuned by the developers of the algorithm and are not changed in many applications. However, following the work by Isensee *et al.* [66], we also vary these fixed parameters here. The rule-based parameters are optimised automatically using heuristics that are extracted using a "dataset fingerprint", calculated for each new dataset the network is being trained on. These parameters include, for example, the network topology, batch size, etc. The empirical parameters must be tuned manually. The overall workflow used to tune and train the nnU-Net is illustrated in Figure 3.

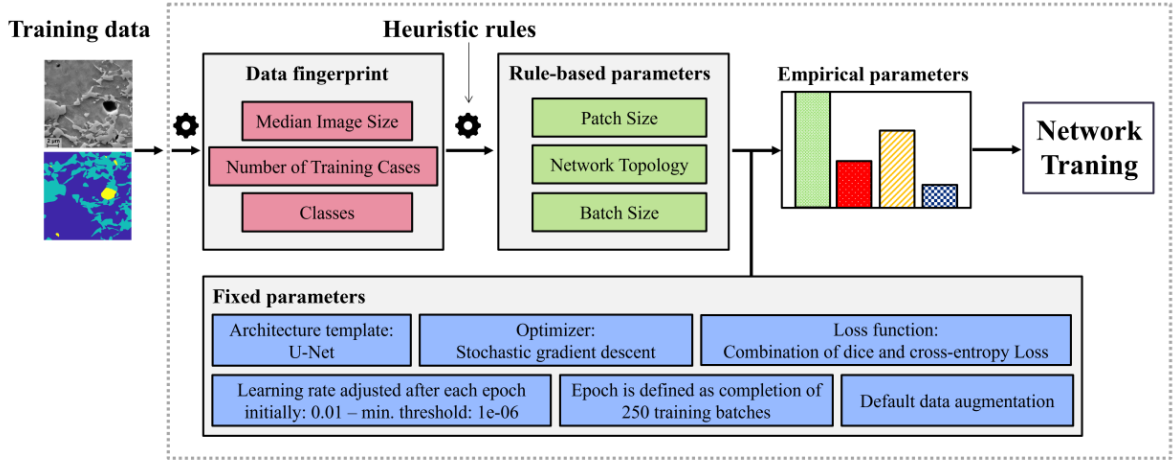

*Figure 3: The adapted and simplified flowchart of the nnU-Net method, based on [66], for the current dataset.*

The default values of the fixed parameters include: the architecture template (UNet), the choice of optimizer (here stochastic gradient descent is employed, following the methodology presented in the original work [62] and the findings of [67]. In this study [67], higher generalizability is observed with stochastic gradient descent in comparison to adaptive methods)the learning rate schedule (starting from 0.01, following $l_r = (1 - \frac{e_p}{e_{p,max}})^{0.9}$ where $e_p$ is the current epoch and $e_{p,max}$ the total number of epochs), the training schedule (1000 epoch are defined as, each epoch completes 250 training batches), the choice of loss function (cross entropy) [68]. The default data augmentation (rotations, scaling, Gaussian noise, Gaussian blur, brightness, contrast, simulation of low resolution, gamma correction and mirroring).

In total, we have 171 large-scale experimental images that we split as described in the following, depending on the training scenario. To be able to test the trained network on an independent dataset that was not seen by any network during training, we set seven large-scale images aside for this purpose. The remaining images were split such that 131 are used for the training of the network (the training dataset) and 33 images for independent evaluation in all cases except for the cross-validation. These 33 images form the validation dataset. In all cases but the cross-validation approach, the images that form the validation and training dataset are chosen randomly before training any of the networks and are kept fixed for all network configurations. For the case of cross-validation, 131 plus 33 images (i.e. the training and validation dataset) are given to the network and the cross-validation procedure then randomly chooses the split between the training and validation data using the same ratio as in the other cases. As mentioned above, the seven images in the test dataset are not seen by any network during training.

Additionally, we investigated the following changes to the fixed parameters:

- **Data split:** We explored different data split strategies, with 5-fold cross validation [66] and fixed validation set, and augmentation of the training data (up to 520 training data). The data augmentation techniques applied here are a translation operation, flip operation, rotation operation of 90° and 160° and a shear operation.
- **Loss function**: Apart from Dice and cross entropy loss (CE) in the original work, following some medical image segmentation approaches, we have explored other loss functions. Here a combination of TopK loss (TopK) and dice loss (Dice) have been explored besides CE [69].

Heuristic rules in correlation to the hardware capabilities (e.g., GPU memory), define the rule-based parameters. These parameters for our data are: batch size (8,12), patch size (median image size, 512 × 512 pixels) and the network topology (like the number of down sampling layers). The different sets of experiments explored while tuning the network parameters are illustrated in Figure 4.

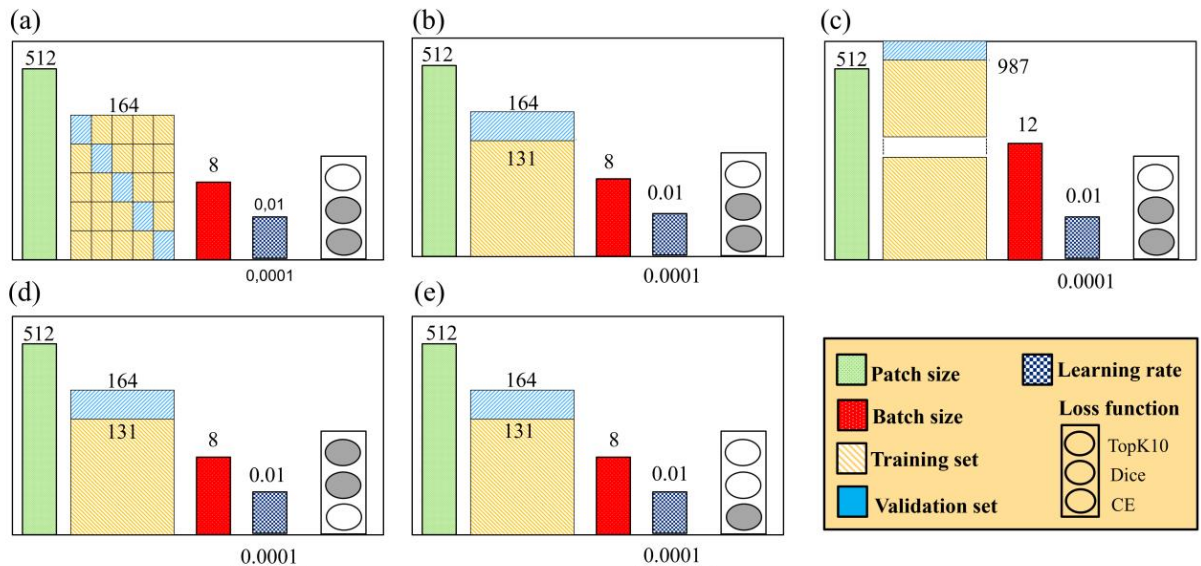

*Figure 4: a) data split: 5-fold cross validation, loss function: Dice + CE, b) fixed validation set , loss function Dice + CE, c) Training data augmentation , loss function : Dice + CE, d) fixed validation set , loss function : TopK + Dice, e) fixed validation set , loss function: CE .*

The process of predicting large images is done by using a sliding window [62]. Specifically, the network rasters across the panoramic image and segments it into windows of the same size as the patch size utilised during the network's training, here $512 \times 512$ pixels. There is an overlap of half the patch size between the neighbouring predictions. The predictions are combined by averaging the softmax outputs of the network across all predictions. The accuracy diminishes towards the edges of individual predictions due to the padding applied in convolutions. To avoid stitching artifacts, Gaussian importance weighting is applied to the resulting prediction, assigning higher weights to the canter voxels during softmax aggregation [62].

Training the network was performed on central high-performance computing facilities of RWTH Aachen university, with a GPU node providing a Nvidia Tesla V100 graphics card, and 16 GB RAM, as well as 1 TB storage.

### 2.1.2 Generative Adversarial Networks

On a high level, GAN [70] consist of two competing networks: One network, the generator, is used to create realistic images, and the other, the discriminator, tries to tell the generated image apart from real images. The training of the generative adversarial network stops when the discriminator network is no longer able to tell the "fake" generated images apart from real images. More formally, GANs learn the mapping from a random noise vector $z$ to some output image $y$: $G: z \rightarrow y$. A variant, the conditional GANs (cGAN) [71], allows conditioning on a quantity of interest in the image generation process. Building on this approach, pix2pix GANs [72] use images for the conditioning, such that this kind of network learns the mapping from an image $x$ and the random noise vector $z$ to some output image $y$: $G: \{x, z\} \rightarrow y$. This approach allows the "translation" of images, such as, for example, turning black-and-white images into colour images. In our case, we can use this to train the network to produce segmented images from the (raw) scanning electron micrographs by providing a set of already segmented images as part of the training process. A generic workflow of image segmentation using this method is shown in Figure 5.

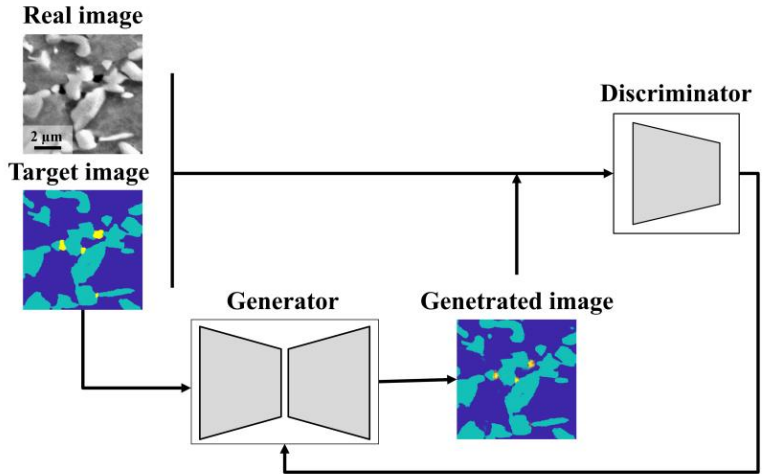

*Figure 5. Generic workflow of image segmentation by the pix2pix GAN.*

Since the pix2pix network uses images with a size of 250 × 250 pixels, the original images were cut into four images each, such that a total of 4 × 171 = 684 images were used for training and testing. We used a batch-size of 10 and a learning rate of 0.0002 during the training process and trained the network for 150 epochs. The choice of the batch size in both methods was a trade off with respect to the hardware capabilities and method performance.

Visually, the output from the pix2pix GAN includes artifacts that mostly affect the ferrite and martensite islands (Figure 6.b). Therefore, we propose and include here the following procedure as a postprocessing step to improve the quality of the segmented image: Similar to the work in [30], we use the DBSCAN clustering algorithm [73] to identify clusters of pixels that can be attributed to damage. Typically, damage appears as an agglomeration of black pixels in the electron micrograph. However, due to the imaging process, the black level varies across all damage sites on the micrograph, meaning that we cannot use a single colour threshold. Therefore, we used the dynamic colour thresholding algorithm Yen [74] to separate the black pixels. Next, the DBSCAN algorithm was applied to detect the islands with black pixels specifically in the microstructure. The use of DBSCAN algorithm often results in the exclusion of very small clusters (noises) of black pixels which do not actually form voids. Nonetheless, some small black voids are detected in this step, for example shadows (Figure 6c). Therefore, in a subsequent step, only the damage sites predicted by both pix2pix GAN and DBSCAN are selected to remain in the final images (Figure 6d). The damage sites identified by both methods (yellow clusters in pix2pix and black clusters in DBSCAN result), represents a proper alternative to actual damage sites. However, the ferrite and martensite masks retain some artefacts like open islands or dispersed pixels of martensite on the ferrite island (some examples shown by orange arrows in Figure 6e). Finally, a morphology correction step, was applied to clean the dispersed noises (cleaning) and a closing operation was applied on top of that to fill the open islands of martensite where some missing pixels exist (Figure 6f). Due to the combination of the methods applied here, we call this method "DB-GAN ensemble" within this work.

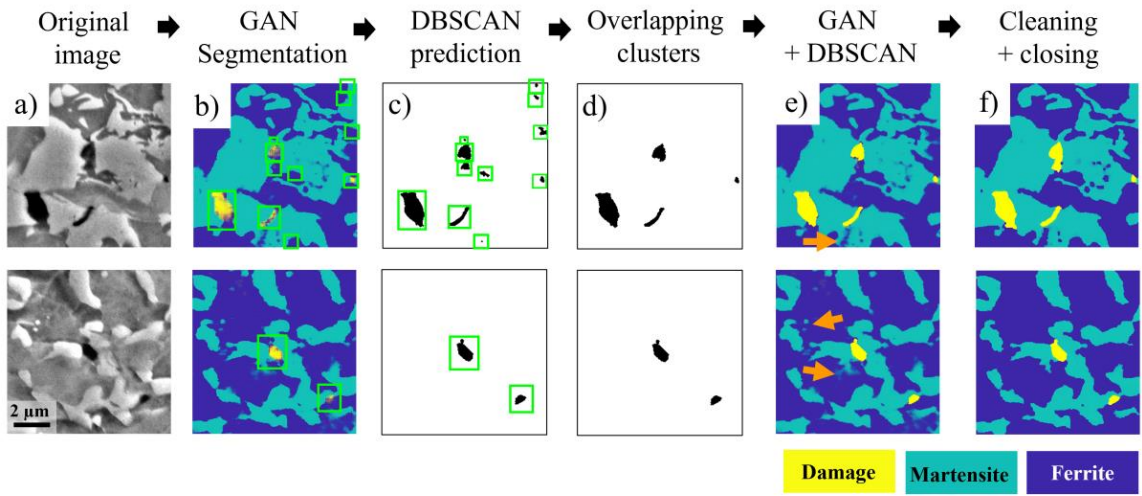

*Figure 6. Step by step representation of the DB-GAN ensemble method. a) Original SEM image, b) segmentation result of pix2pix GAN, c) the damage sites identified by the DBSCAN clustering algorithm, d) the damage sites identified by both DBSCAN and pix2pix GAN, e) overlay the results of pix2pix and DBSCAN, f) the result after the morphological corrections by closing and cleaning. Orange arrows indicate the artifacts of pix2pix GAN.*

Due to the internals of the GAN-based network architecture, the segmented images are limited to a size of 256 × 256 pixels. Therefore, we need to crop the large-scale panoramic images to smaller patches of this size, run the segmentation and postprocessing procedure, and then stitch the images back together. An example of this process is shown in Figure 7. This additional cutting and stitching step can introduce additional artefacts that are not present in the nnU-Net-based approach as the latter can handle much larger images by the sliding window approach. In principle, a sliding window approach could also be implemented for the GAN-based analysis. However, as this requires not only the method itself but also rules for dealing with differences in the overlapping areas, we have refrained from implementing this additional step with a few to avoiding additional and more difficult to interpret artefacts.

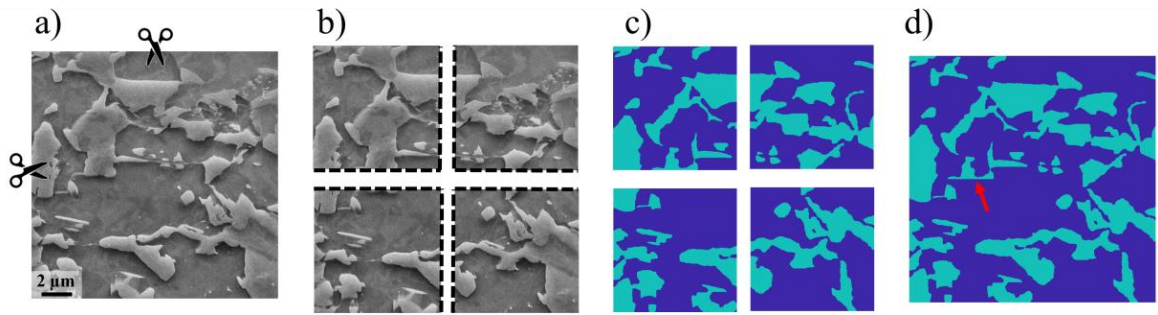

*Figure 7. a) SEM image with 512 × 512 pixel size, b) 512 × 512 pixel images cropped to the size of the training data, c) application of the DB-GAN ensemble method, d) stitched, segmented image. The red arrow indicates the error introduced due to cropping and stiching.*

## 2.2 Material and experimental methods

Damage in dual phase steels occurs as a result of different deformation mechanisms at the microscale within and between the two constituent phases [51, 75]. The propensity of dual-phase (DP) steels to incur damage and failure during quasi-static testing is strongly correlated to the local microstructural characteristics giving rise to local internal strains [76, 77], and has a pronounced impact on the fracture toughness [78-83] and fatigue crack growth resistance [79, 84, 85]. Important microstructural features therefore include those associated with the properties of the individual phases, such as the martensite carbon concentration [86], martensite internal structure [87], hardness of martensite [88], and strain hardening capacity of the ferrite [89], or their morphology within the dual phase microstructure, characterised by the volume fraction of the martensite phase [86, 90-92], and the morphology of the ferrite grains and martensite islands [93-95], the microstructural morphology, the grains' or islands' size and aspect ratio, and the interconnectivity [96]. Furthermore, the nucleation and evolution of deformation-induced damage naturally depend on the amount and direction of applied strain, and any changes in the strain path [30, 37, 38]. These in turn give rise to the heterogeneous local strain distribution originating from the two phases' mechanical contrast, as revealed by high-resolution microscopic-digital image correlation (μDIC) and crystal plasticity (CP) simulation.

The strong interest in dual-phase steels has already attracted considerable interest in microstructure analysis [97-100], particularly in terms of phase segmentation. EBSD-based approaches are typically challenging for this kind of material, since the crystallographic relationship between the two phases (ferrite and martensite phase with BCC and BCT crystal structures, with small tetragonality in the martensite phase) is very close, and phase separation becomes even more challenging where the internal strain is high, i.e. particularly after deformation. This can be overcome, to some extent, by combining several approaches, for example, electron channelling imaging (ECCI) [89, 101], electron probe microanalysis and transmission electron microscopy (TEM) [102, 103]. Recent work combined orientation information collected from EBSD and its associated kernel average misorientation (KAM) and pattern quality maps using a UNet segmentation algorithm [45], however, the approach lacks decent transferability from the pristine, undeformed material to deformed samples with high internal strains.

The most accurate of the above methods, i.e. those involving ECCI and TEM and to some extent also EBSD-methods requiring very high pattern quality, are confined to relatively small areas. In contrast, the large variability of the microstructure mandates the extraction of phase assignment at sub-micron level of individual martensite islands and damage sites, but across large scales approaching the sheet thickness to capture the commonly encountered martensite banding. High resolution, segmented data across large areas of the order of at least 1 mm² is therefore essential to be able to represent the microstructure in subsequent analyses or modelling in a meaningful way. Here, we present a method that is capable of segmenting panoramic images acquired in the SEM that resolve sub-micron features across an area of the order of a mm² without requiring diffraction information or conditions or clear grey-scale separation of the two phases after etching.

To image and segment a common and at the same time challenging microstructure, we used a commercial DP800 dual-phase steel (ThyssenKrupp Steel Europe AG) with a tensile strength of the order of 800 MPa. The chemical composition and mechanical properties of this steel can be found in [104]. Samples were cut out of a 1.5 mm thick sheet metal into dog bone shaped tensile specimens and then elongated uniaxially in the rolling direction specified by the manufacturer. The tensile tests were performed using a microtensile stage (Proxima 100; MicroMecha SAS, France) with a gauge length of the tensile samples of 3.65 mm and a square cross-section of 1.5 mm side length. After testing, the samples were cut to the gauge length with an electron discharge machine.

The initial surface was prepared by grinding the surface manually using sandpapers starting from 800 to 4000 grit. Afterwards, the samples were polished mechanically with 6 μm, 3 μm, and 1 μm diamond suspension with an alcohol-based lubricant (99.5% Ethanol + 0.5% Polyethylene Glycol), and final polishing and cleaning was performed using colloidal silica suspension (OPS). Finally, each sample was subjected to a 1% Nital solution for 10 seconds, leading to a visible phase contrast between ferrite and martensite in the electron microscope due to preferential etching of the ferrite phase. A typical microstructure of this steel after etching is shown in

Figure 9 in plane-view and cross-sectional view. Typically, martensite islands are visible in light grey, while the ferrite matrix is darker grey, and deformation-induced damage appears as voids in black. The banded martensite is visible directly in the sheet cross-section (

Figure 9a), and indirectly in the variable martensite density in plane-view. Typical deformation induced damage sites [38] are shown inset at higher magnification.

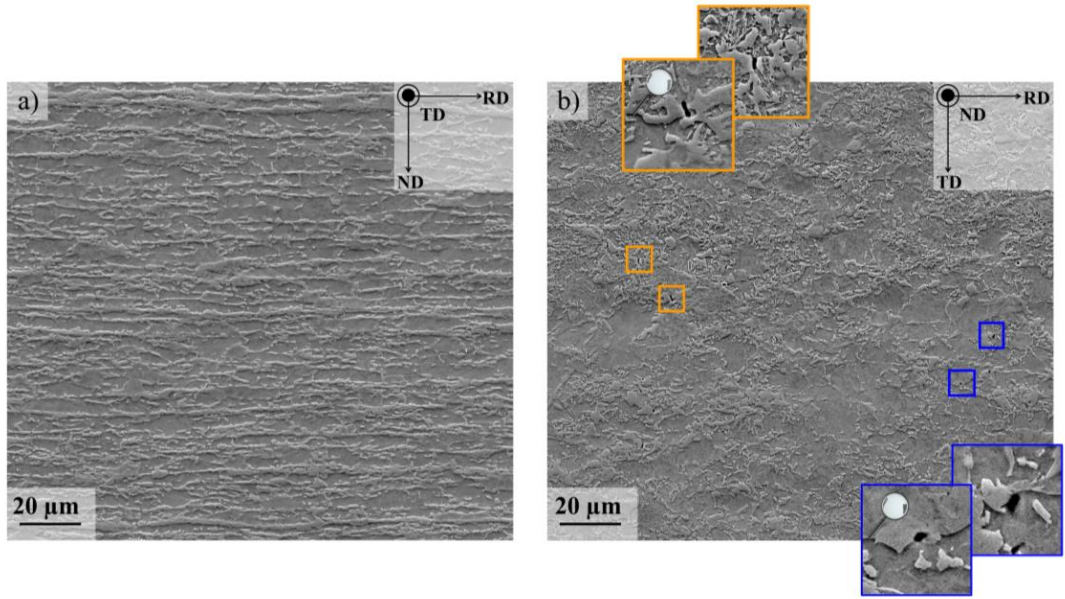

*Figure 8. Micrsotructure of the DP800 steel: a) cross-sectional view revealing banded martensite and b) in plane-view with magnified examples of the two main mechanisms of deformation-induced damage inset (orange: martensite crack, blue: interface decohesion).*

# 3 Results
## 3.1 nnU-Net

The resulting network topology for the corresponding SEM images of the DP800 dual phase steel is illustrated in

Figure **9**. After the automatic optimisation of the rule-based parameters using the dataset fingerprint, the final nnU-Net topology consists of seven down-scaling operations, and the network is initialized by 32 feature maps at the first place and then increasing to 480 at last. Finally, the down sampling operation is terminated when the feature map results in size 4 (at the bottom row) continued symmetrically by up-scaling operations.

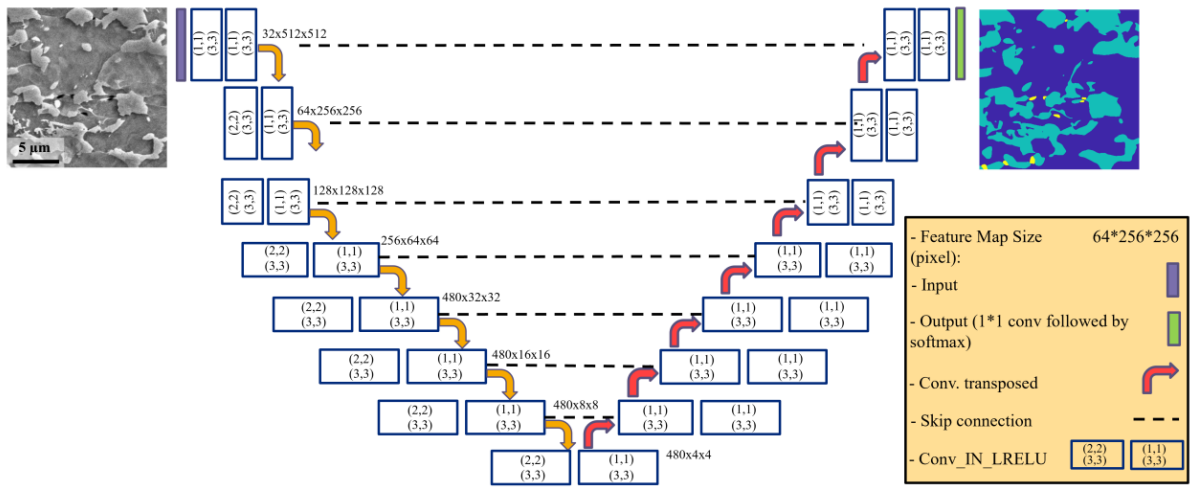

*Figure 9. The optimized U-Net topology for the data*

From the different packages of experiments, the package with a fixed validation set and with (CE + Dice) loss function was selected. This package achieved 96.2% and 82.9% for validation accuracy and dice score respectively on the test set, (marked with red box in Figure 10). The evaluation metrics of all packages of conducted experiments and the respective values for the individual classes for the selected package are also given in Figure 10. The validation dataset refers to the average of the splits into training/validation data chosen automatically during the cross-validation approach, and, for the other cases, to the fixed validation dataset consisting of 33 images chosen randomly before the training of any package starts.

The quantitative evaluation of the results, based on pixel-by-pixel comparison, did not show significant differences for the various packages. However, the package associated with a fixed validation set (highlighted by the red rectangle in Figure 10) exhibited the best visual performance on both the test and validation sets, according to the judgment of material scientists. In particular, better visual performance judged by human experts involved replication of fine details in the microstructure, such as thin protruding pixels from larger islands or thin connecting segments between islands. Although here the images chosen for the validation set are arbitrary, opting for the package with 5-fold cross-validation might be a prudent choice to try in other experiments, due to the mixed nature of the data split, favouring unbiasedness.

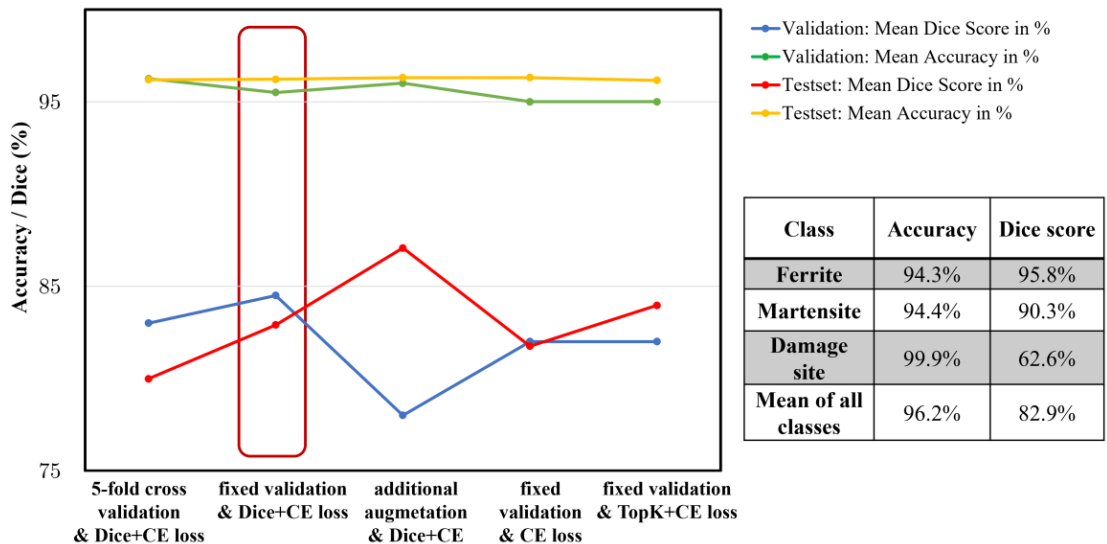

*Figure 10. Evaluation metrics. Accuracy and dice score for the validation and test set for all packages of experiments. The selected package (marked by the red box) possessed the best visual performance. For this case, the related values for each class are also shown in the inset table.*

## 3.2 DB-GAN ensemble

For the evaluation of the microstructural images segmented by the GAN and DB-GAN ensemble method, we calculated the average confusion matrix over ten confusion matrices from ten randomly selected test 250 × 250 images. These images were randomly cropped from 512 × 512-pixel images that were excluded from the training data. The segmented images generated by pix2pix GAN and DB-GAN were compared with manually labelled data. Specifically, each pixel in the segmented images was compared with its counterpart in the manually labelled image. The results are depicted in Figure 11. As shown in this figure, the performance of the GAN method was much improved following the addition of DBSCAN clustering and morphology correction.

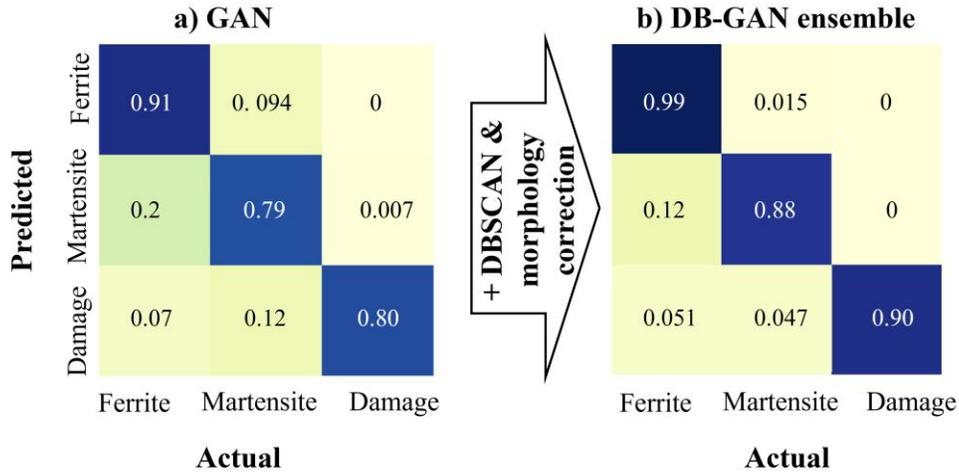

*Figure 11. The average confusion matrix, comparing the performance of the pix2pix GAN and DB-GAN ensemble for ten randomly selected 250 × 250 pixel images.*

## 3.3 Comparison of the nnU-Net and DG-GAN ensemble methods

To compare the performance of the DB-GAN ensemble method with the nnU-Net method, we extracted the confusion matrices for ten randomly selected test images that were excluded from the training data but with almost the size of the pix2pix GAN training data (250 × 250 px). The results are shown in Figure 12. We observe that both methods exhibit comparably reliable performance.

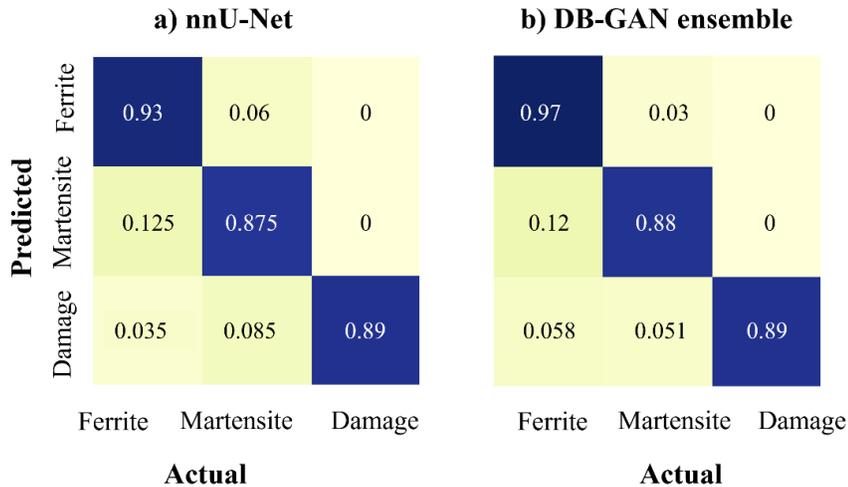

*Figure 12. The average confusion matrix, comparing the performance of the nnU-Net and DB-GAN ensemble for randomly selected 250 × 250 pixel images images.*

# 4 Discussion

Using the example of a commercial DP800 dual-phase steel, we have demonstrated two approaches based on artificial intelligence to segment large panoramic electron micrographs. Here, we discuss first the performance and applicability of the two approaches before highlighting how the information obtained by analysis of such large, stitched micrographs or micrograph arrays can be used for further analyses and investigations.

## 4.1 Comparison of Image Segmentation Methods
A key difference between the two approaches lies in the use of fixed input size image segments of 256 × 256 px in case of the DB-GAN ensemble method versus direct application to full scale panoramic images of the nnU-Net approach.

### 4.1.1 Side-by-side comparison of individual features
Figure 13 shows a side-by-side comparison of the two approaches for 512 × 512 px SEM images with the corresponding ground truth data from manual labelling.

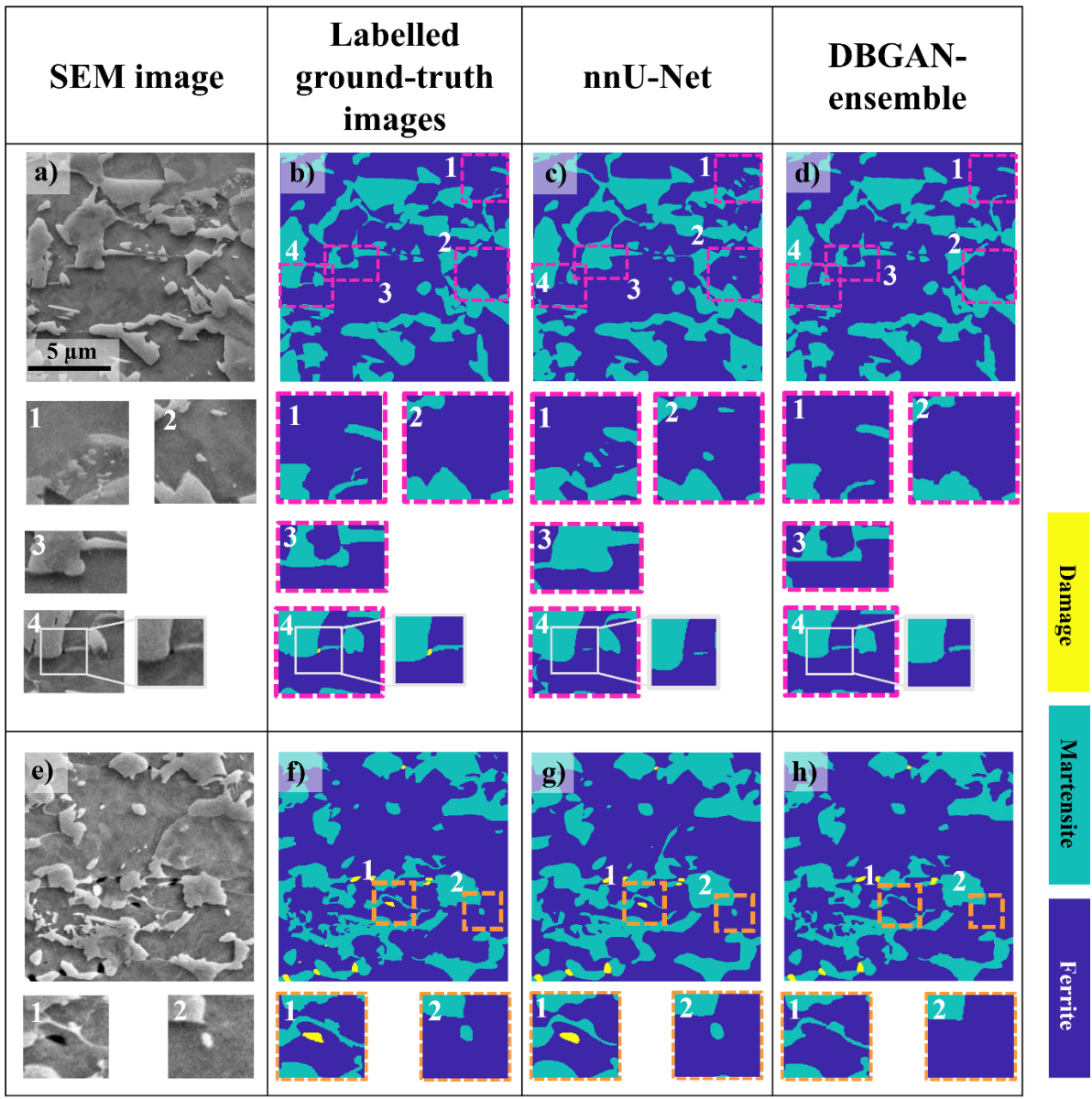

*Figure 13. Examples of SEM images along with the manual labeling, ground truth masks. Side by side comparsion of the images segmented by nnU-Net and DB-GAN ensemble method.*

Overall, either approach yields acceptable results, as would be expected based on the similar performance metrics. However, examining the segmented images in comparison to the original SEM images (areas 1 and 2 in Figure 13c) and Figure 13d), nnU-Net achieves a higher precision pixel-by-pixel compared to the DB-GAN ensemble approach and is able to preserve finer details of the microstructure in the segmented image. Some of these details can even be related to the remainder of the OPS polishing solution that has not been introduced to the classes and neither labelled in the ground-truth data as it is typically not considered in the analysis of the microstructure. An example of such areas is shown in Figure 14 with b) and c) showing two different, manually labelled segmentations of the same microstructure. In Figure 14b), the mask in the ground truth data is not labelled with a pixelwise precision, and Figure 14c) shows an additional mask applied with a pixelwise precision. This is a typical case where the subjectivity of manually labelled data can introduce uncertainty to the evaluation metrics and the real results. It is evident that the labelling process itself is a source of uncertainty as different experts will, in general, attribute pixels of the intricate features of the microstructure differently to the phases. Nonetheless, some of the fine details have been segmented by the nnU-Net method in Figure 13c) area 1 and 2. Where the white OPS particles have been segmented, they were identified and segmented as martensite islands. This is unsurprising given that they share the closest colour resemblance to the martensite islands and a separation of the two is challenging for the human eye as well, particularly where agglomerates preclude identification based on spherical shape, as is the case for the largest such particle in Figure 14.

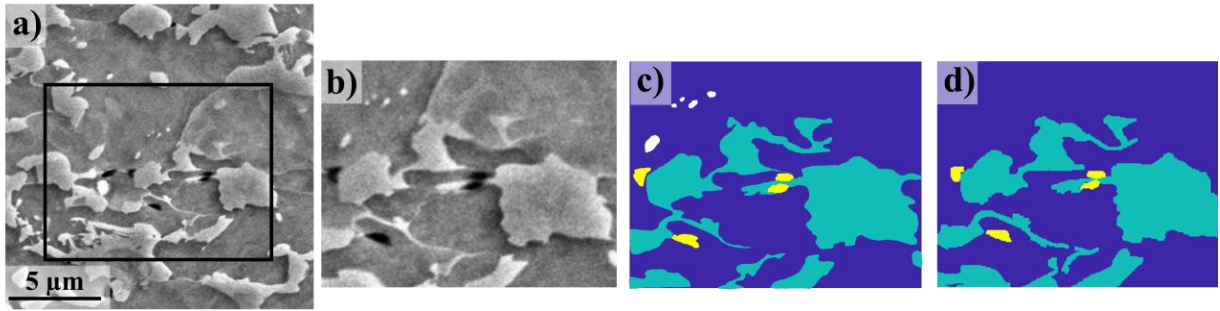

*Figure 14. a) SEM image of the microstructure and two different, manually labelled versions of the same image area. b) Manual labelling taken directly from the ground truth dataset, c) area relabelled with higher, pixelwise definition of the existing phases and OPS particle as an additional fourth class.*

Similarly, a damage site and a small white island have not been identified by the GAN-based segmentation algorithm (areas 1 and 2 in Figure 13h), in contrast to the nnU-Net based segmentation (areas 1 and 2 in Figure 13 g). Comparing the areas 3 of Figure 13c) and Figure 13d), the stitching line artifact is evident for the DB-GAN segmented image, whereas this is not the case for the nnU-Net segmentation. In area 4 of Figure 13c) and Figure 13d), both methods have missed the detection of the small damage site, with size around 2 pixels (50 nm), marked by a tiny yellow dot in the ground truth data.

### 4.1.2 Generalizability

A key aspect of machine learning approaches is to investigate their performance on out-of-distribution samples. In particular, we are interested in the performance of the trained networks on new data that does not originate from, for example, the exact same experimental specimen or material that was used to obtain the training data from. In our case, we have applied the trained networks, both for the nnU-Net, as well as the DB-GAN ensemble method, to electron micrographs of the DP800 steel that was subjected to a heat treatment, which changes the microstructure. Compared to the original sample, the martensite islands are finer and more circular, as opposed to more elongated and banded martensite islands in the as-received DP800 sample. A side-by-side comparison of the two microstructures is shown in Figure 15 with the micrograph in a) taken of the original commercial sheet material that was used to train the networks and the micrograph in b) showing the microstructure after a heat treatment of the original material at 775 °C for 330 seconds followed by water quenching for martensite formation.

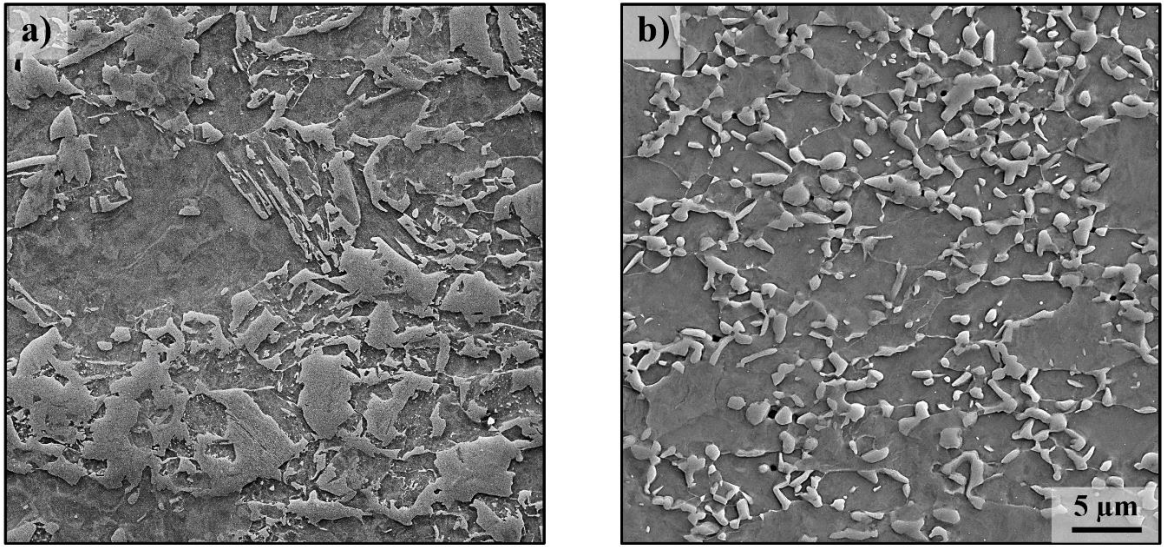

*Figure 15. Comparison of the microstructure configuration of a) the as-received dual phase steel and b) the same steel after heat treatment (775 °C for 330 seconds followed by water quenching).*

In Figure 16, we show examples of how either method performs when applied to images obtained from a sample that was not used during network training, i.e. from the new microstructure shown in Figure 15b. Both methods work well on this previously unseen but visually similar image data. However, again some deficiencies can be seen for both methods on the finest details. If we consider the precision in the detection of the damage sites, the nnU-Net performs better than the DB-GAN ensemble in Figure 16a) and b). On the other hand, Figure 16c) and d) show two examples where the GAN-based approach yields better results, whereas nnU-Net fails to identify the complete area of the damage site. In summary, both methods perform reliably on new microstructural configurations, however both methods can fail to yield perfect results, in particular when there are relatively fine details in the microstructure. As it requires lots of manual labelling, there is no quantitative comparison for the representation of fine details. It is important to keep in mind, however, that while either method show some challenges in processing the finest details of the microstructure perfectly, human experts would likely show the same, if not worse outcomes when processing large-area micrographs of several hundreds of micros in x- and y- direction.

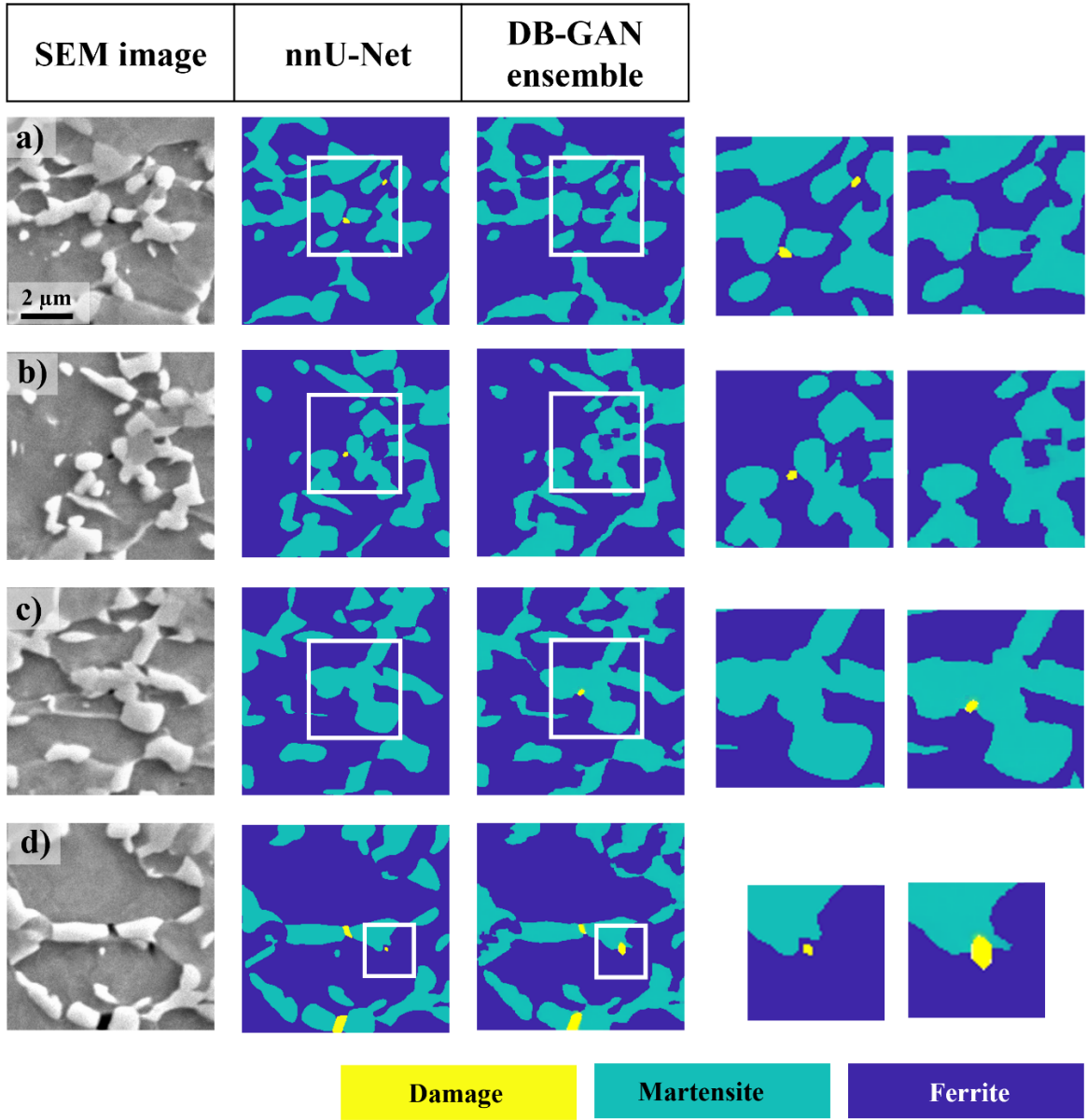

*Figure 16. Comparison of the performance of 2 methods for heat treated dual phase steel (not included within the training data). Images with size equal to training data of DB-GAN ensemble method.*

### 4.1.3 Usability for microscopists

Overall, either approach can be used for image segmentation, provides accurate results, and generalises well to data not seen during training of the machine learning algorithm. The approach based on nnU-Net is almost fully automated and can be applied as a standardised procedure to new scenarios. Compared to the GAN-based approach, it can also be applied to larger images, thus avoiding the additional step of cutting and stitching the electron micrographs as part of the procedure. Additionally, further image artefacts due to the stitching of the GAN-segmented images can be avoided.

The DB-GAN ensemble method is much less automated at present and requires several steps, such as, image pre- and postprocessing, cropping and stitching, and potentially correction of associated errors. These extra steps need to be implemented and controlled by the user. The whole process may, therefore, be less convenient for a user interested in fully automated segmentation. However, in some cases the explicit control over individual steps of the segmentation process may be advantageous. We anticipate this to be the case where colour thresholding is of importance. In our case, this was shown to apply where a focus is put on resolving the damage sites accurately, rather than the segmentation of the ferrite and martensite phases that naturally contribute the majority of the imaged surface area. Such intervention in the learning process is more difficult in the nnU-Net method, where changes in performance of the segmentation due to adjusted colour contrast of the input images may equally be expected but iteration towards an improved pre-processing procedure will be much less straightforward.

## 4.2 Application of large area segmentation 1: Phase fractions in an inhomogeneous dual phase microstructure

A key quantitative parameter in microstructure characterisation is the phase fraction of the constituent phases. The dual phase steel used here as an example, with martensite bands undulating through the sheet metal, presents a particularly challenging case for this, as phase segmentation by simple thresholding is difficult, while manual selection and labelling of small sections can lead to a large variance in the results. In this case, the result will depend strongly on the exact position and size of the windows from which the area fraction was determined.

Using the large-scale segmented images, we can calculate the phase fraction based on a suitably large area, such that the local variations of the microstructure no longer play such a prevalent role. We illustrate this by the following procedure: Choosing nine different window sizes, ranging from approximately 400 $\mu m^2$ to 108900 $\mu m^2$ (equivalent side lengths are given in Figure 18), the martensite phase fraction is calculated in each of these windows. We used two different approaches to distribute the windows across the panoramic image: random placement across the panoramic image stipulating zero overlap and placement around nine regularly spaced points of a 3 × 3 grid. The latter allowed us to compare measurements using the entire area of the square panoramic image, while the first gives a more realistic scenario for measurements of area fractions without the use of panoramic imaging. Both approaches are illustrated for selected window sizes in Figure 17. Random placement was employed up to window size 8 as displayed in Figure 18.

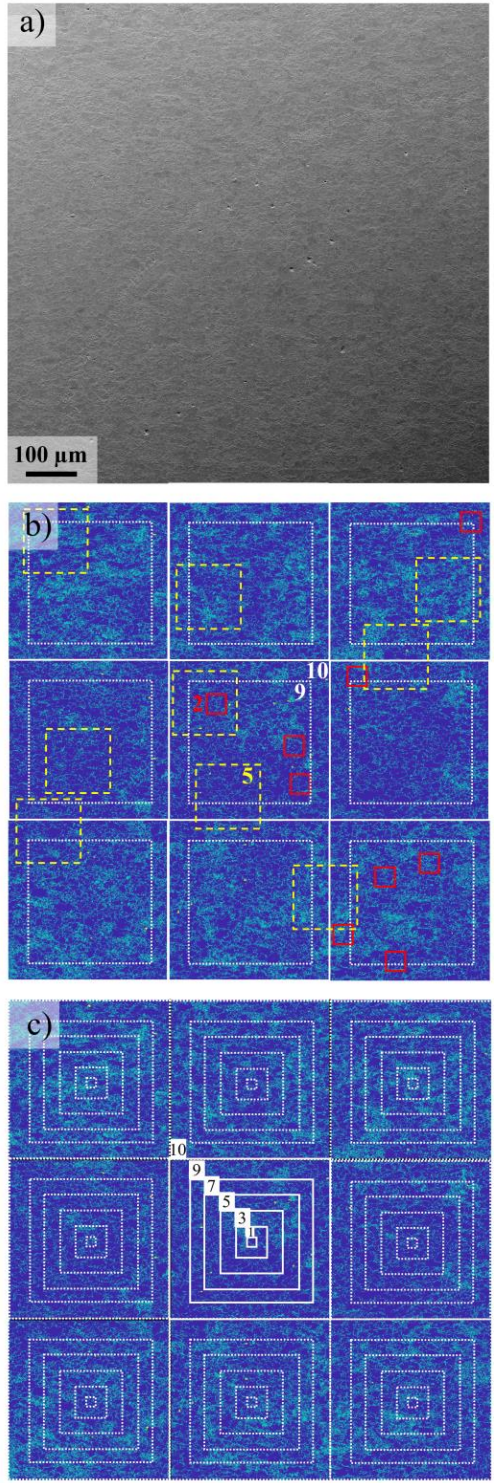

*Figure 17. a) panoramic SEM image and segmentation of the image shown in (a). Bounding boxes of different size were used for phase fraction calculation and replicated 9 times across the panoramic image either (b) randomly (red and yellow dashed squares) or in b) a regular grid to achieve maximum area without overlap. The numbers given correspond to the sizes given in Figure 18.*

From these boxes, we calculated the mean and the respective standard deviations as well for each group of nine windows of equal size. The result is shown in Figure 18. We can clearly see that the smaller windows show a very large variation, as expected, as these measurements are much more sensitive to the local variation of the microstructure, compared to larger windows. This result exemplifies the importance of large and multiarea analysis for quantifying the widely heterogeneous microstructures like the DP800 steel in the current research. The results confirm the possibility of a high fluctuation of the results even for reasonably large areas. This is highlighted here in Figure 18a) when comparing the measurements for sizes 3 and 4, which may look small considering the illustration in Figure 17 but correspond to a total measured area of a square with a side length of 183 and 294 μm, respectively. In addition, despite the observed convergence in the determined average martensite phase fraction by increasing the analysed window size, a noticeable dispersion in the values persist at the window size of 330 μm × 330 μm. We note that this represents a large area equal to 1 mm² in total for all 9 measurements and therefore would likely be uncommon in past literature for materials such as this DP800 steel, which contains phase detail to the sub-μm scale but cannot be reliably segmented using thresholding across a large area or by using constant settings across many different images. The persistent scatter is therefore likely due to the even larger length-scales of the undulating martensite bands intersecting the surface. A comparison between the size-dependent data for randomly placed windows for phase fraction analysis (sizes 1-8 in Figure 18a) and the same analysis for regularly arranged, concentric windows (Figure 18b), underscores this assumption. In the latter case, the distribution of values for each window size appears self-similar and even for the largest window size, each measurement is clearly not representative of the microstructure as a whole. Phase fractions are commonly given as whole %, but without measurements across even larger areas, the minimum size for a reliable measurement that discriminates phase fractions reliably to within 1 % cannot be calculated. In the future, automated panoramic imaging coupled with large area phase segmentation, as presented here, may however help provide quantifications of microstructure variability to be used instead, e.g. by measuring fluctuations along long line-like scans (instead of the square shape used here) across a surface. In any case our results highlight the fact that martensite phase fraction statements from area measurements using microscopy in dual phase steels may be associated with very large errors that sensitively depend on the underlying area that was evaluated and how this area or segments of the total area measured were placed with respect to long-scale microstructure fluctuations, such as the martensite bands in our dual phase steel. Although this result would appear entirely expected, at least with respect to the deviations between measurements of smaller areas, the majority of publications only state how a phase fraction was measured. They thereby omit the essential information on the total area investigated for this purpose, in spite of the uncertainty of the measurement, which we show here may commonly exceed the accuracy at which the phase fraction measurements are stated and compared.

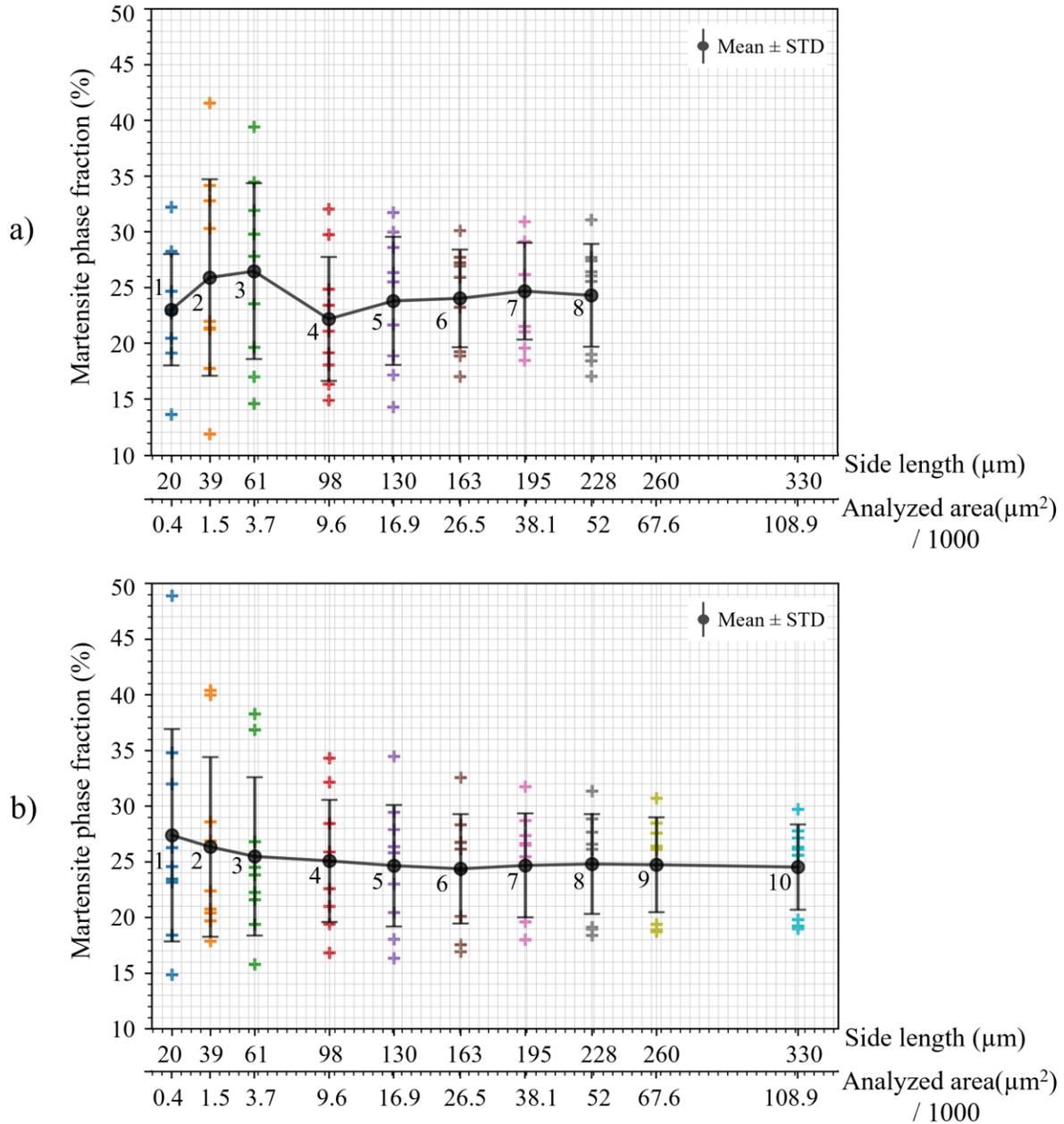

*Figure 18. Area fraction of the martensite phase from 10 different window sizes and using either random (a) or regular (b) spacing of the windows (cf. Figure 17). Each area fraction is calculated from 9 windows, for which individual values, average and standard deviation are given.*

## 4.3 Application of large area segmentation 2: Microstructure data for modelling

Finite element simulations can be used to analyse the properties of a given microstructure numerically, such as, for example, to derive a stress or strain map of the material. However, as a prerequisite, one either needs to employ measures that allow the creation of meaningful synthetic microstructures or, alternatively, "convert" experimental microstructural data into a suitable numerical representation. Here, we again exploit the capabilities of the nnU-Net segmentation algorithm to analyse a large panoramic electron micrograph, which allows us to either model a large area or extract many smaller simulation sites for the same experimental sample and conditions featuring specifically or randomly selected sites, as illustrated in Figure 19.

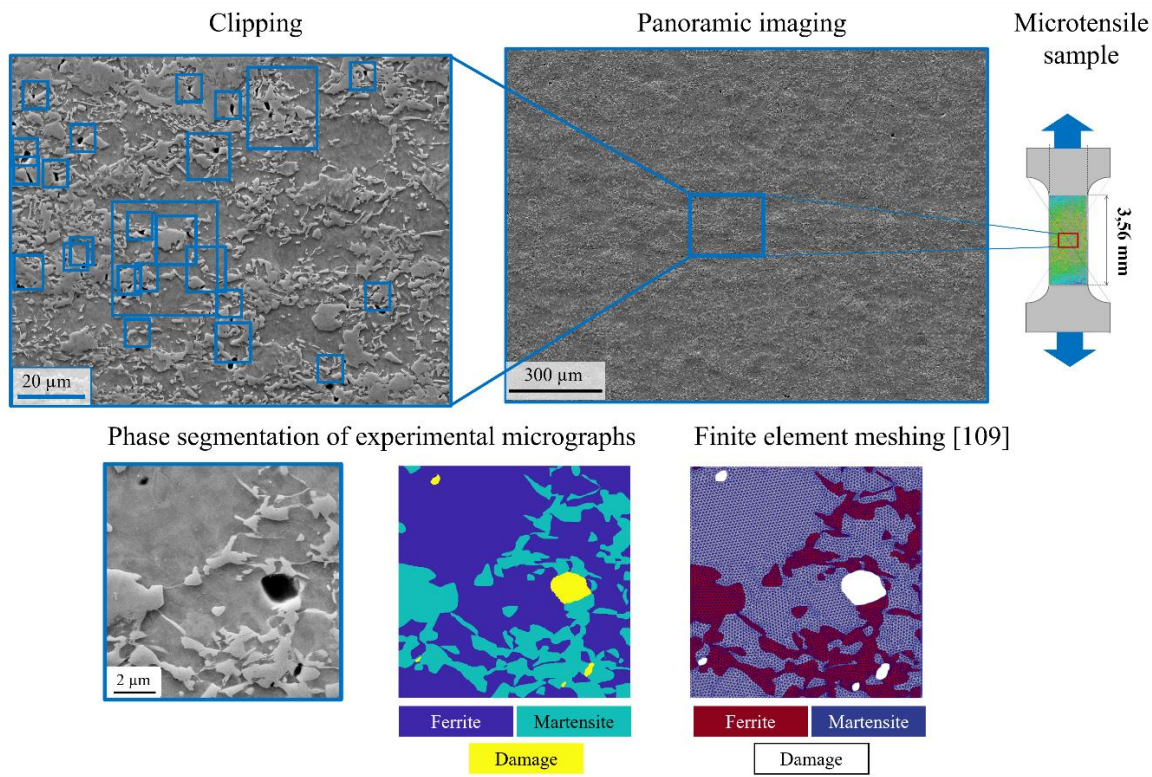

*Figure 19. Illustration of the use of segmented images for finite element modelling of local stress or strain fields. FE meshing taken from [105], published under a Creative Commons CC BY license.*

Segmented microstructural images from large areas of a sample enable a free choice of simulation cell size to capture the desired features and processes at small scales, e.g. damage nucleation at individual phase boundaries, and also allow scaling on the same underlying experimental data to capture phenomena at larger scale, such as the coalescence of many damage sites [106].

## 4.4 Application of large area segmentation 3: Geometric analysis of deformation-induced damage features

Using panoramic images of the microstructure, we are able to study the microstructure across large areas with respect to rare or variable features. In the case of the strained dual phase steel, we consider here the geometry of damage sites. As an alternative to performing direct simulations on the segmented microstructures, as exemplified above, average measures of their morphology can also be used to create synthetic microstructures, inform continuum models requiring damage distribution, size or aspect ratio as input, or to assess the spatial distribution of particular types of damage across a sample or in relation to other microstructural features. Here, we will exemplify two applications, namely the investigation of the arrangement of the phases surrounding particular types of damage sites originating from either martensite cracking or interface decohesion and the geometry of these two different classes of damage in terms of their inclination angle to the major strain axis and the void area. For identification of the underlying damage mechanism at the many damage sites contained in a panoramic image, we employed the convolutional neural networks introduced in [30, 37] using the unsegmented image data for damage type classification (as the neural networks were not trained on segmented data) and continuing on the same sites using the additional segmentation information obtained here.

Figure 20 shows an analysis of the angular distribution of the ferrite and martensite phases around interface decohesion (IF) and martensite cracking (MC) sites from around 650 images of each type. This distribution also represents a major aspect cited by human researchers in their decision to classify image data of damage sites in dual phase steel [30]. With large area segmentation data now available and encompassing many sites, this kind of data can now be used to formulate a representative geometric signature of the two damage types, in order to circumvent the use of neural networks for damage site classification when using segmented data. Conversely, this analysis may be used to monitor (training) data quality as change in this type of analysis may be expected if different classes of damage sites are inadvertently mixed or if data from very dissimilar microstructures is added. In addition, the geometric analysis may also be of use to introduce a sub-division of a class that was not applied during classification network training. Here, the geometric analysis clearly distinguishes partial and through-thickness cracks, as shown in Figure 20c and d, respectively. Such a sub-division may, for example, be useful when following crack propagation during in-situ tests as crack nucleation sites that may allow the observation of crack propagation could be easily identified.

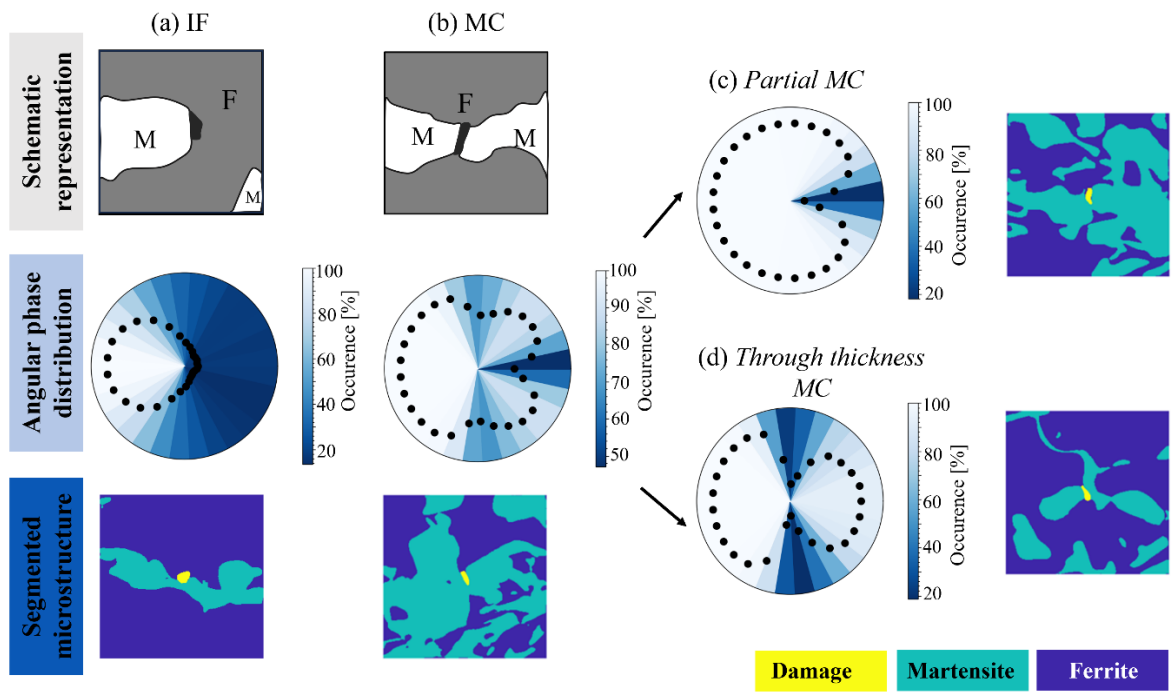

*Figure 20. Angular profile of martensite phase distribution around different types of damage sites. a) Interface decohesion (IF), b) martensite crack (MC). Martensite crack sites comprise two types of cracks partial (c) and through-thickness (d).*

The extraction of geometrical properties related to the damage sites, specifically their inclination angle concerning the uniaxial tensile direction, as well as their aspect ratio and the thickness of the martensite area surrounding them, is presented in Figure 21a (for the methods employed to extract these parameters please refer to the detailed descriptions in the supplementary materials). Although both classes appear to be compatible with a Gaussian distribution, it is evident, that the interface decohesion has a wider distribution and that the martensite cracks are predominantly inclined perpendicular to the loading direction, whereas interface decohesion mechanism does not display a discernible trend. This is in line with the normal stress driven brittle failure of martensite islands and a measure which may change as the stress-state is varied, for example towards biaxial straining [37]. Also consistent with the underlying deformation mechanism, the martensite cracks have a higher tendency to larger aspect ratios, that is a large crack length coupled with smaller crack opening, compared to interface decohesion, which is driven by plastic flow inside the ferrite phase Figure 21b.

In addition to these fairly obvious observations that may be of use where different strain paths or microstructure morphologies are to be compared, the segmented data may also in the future be used to learn more about an individual mechanism and its relation to the surrounding microstructure. As example, we show here an analysis of the surrounding phase's thickness. Again, the distinction between interface decohesion and martensite cracking reveals the expected tendency, namely that in the case of interface decohesion the martensite island is distributed mainly to one size (cf. Figure 20) and therefore a larger thickness of martensite is also found towards one side of the crack Figure 21c. Similarly, martensite cracking is likely to occur at the narrowest point of martensite islands giving a favourable crack path normal to the tensile axis. This is indeed found for the martensite crack here and corresponds to the central minimum of the curve in Figure 21d. Note that the shaded areas in Figure 21c-e indicate the standard deviation of the data around the average martensite island width marked with the full line. This implies that for most of the martensite cracks, a local minimum width can be identified at the crack site with most curves showing a shape similar to the average highlighted in a darker shade in Figure 21c-e. Interestingly, there is again a visible separation between the partial and through-thickness cracks, as plotted in Figure 21e. The larger thickness reduction around the crack in through-thickness cracks may be the result of martensite plasticity during crack propagation or a pre-existing, stronger constriction of the island favouring crack propagation through the entire thickness of the island. Island thickness may also play a role in addition to any stress concentrating features, as revealed by a tendency of through-cracks to be found in thinner islands, consistent with accelerated crack growth as the process zone of the crack exceeds the fracturing volume. While we cannot draw definite conclusions on this matter based on the data presented here, a careful geometric analysis of martensite island shape and any deformation-induced changes in un-deformed and deformed dual phase steel may directly contribute to current research questions surrounding martensite plasticity in dual phase steels [107-109].

Overall, we show here that segmentation may play an important role in assessing microstructural features and interpret their occurrence and correlation with other physical mechanisms like deformation-induced damage formation. Machine learning based image segmentation of large panoramic data may now also serve as an alternative approach to side-step detection and classification of microstructural features, such as damage sites of different types [30], which was not possible previously based on the microscopy data alone. Conversely, where machine learning algorithms for classification of microstructural features have already progressed to high maturity or large training datasets have been generated, subsequent segmentation may still be of use to monitor changing training data sets where a hypothesis about feature shape or the distribution of surrounding phases can be formulated.

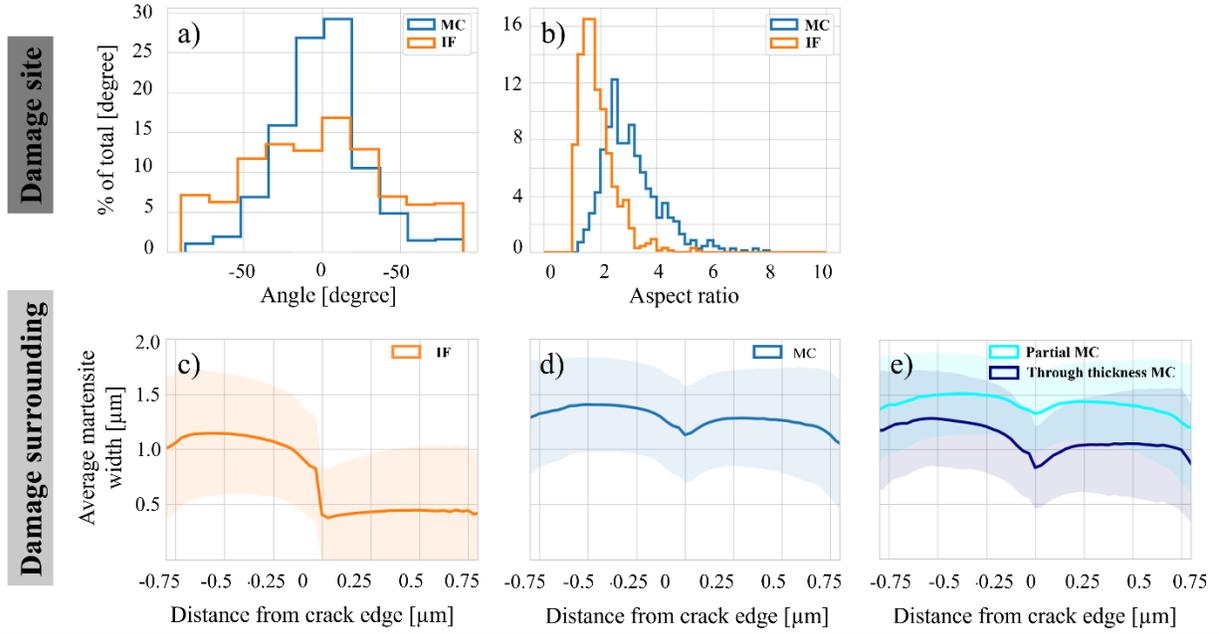

*Figure 21. Distribution of (a) the inclination angle and (b) aspect ratio of two common damage site types (MC: martensite cracking, IF: interface decohesion). The average width of the surrounding martensite for (c) IF and (d) MC sites is shown in terms of the distance from the site/crack edge with the average plotted as full line and the envelope of the corresponding standard deviation highlighted in a lighter shade. In the case of martensite cracks, the data shown in (d) may again be divided into partial and through-thickness cracks (e).*

# 5 Conclusions

In this work, we have introduced two new methods for microstructure segmentation based on artificial intelligence with a focus on applying these to large-scale panoramic electron micrographs. As a case-study, we used a commercial dual-phase steel that is used in a wide range of applications but challenging to segment. Both methods show good performance and can be used for segmenting panoramic electron microscope images. A detailed side-by-side comparison of the methods and their applications lead to the following conclusions:

 i. Both the nnU-Net and DB-GAN ensemble methods exhibit effective segmentation capabilities for electron micrographs on a large scale, showcasing a high degree of generalizability.
 ii. At similar image resolutions, both methods demonstrate comparable pixel-by-pixel accuracy for images of size 512 × 512 pixels (training data for the nnU-Net method) and 256 × 256 pixels (training data for the DB-GAN ensemble method).
 iii. Where a high degree of in-built automation is of essence, the nnU-Net method is preferred, as it segments panoramic images directly without the need for the user to crop or stitch panoramic data.
 iv. In contrast, the DB-GAN ensemble method involves external cropping and stitching. This comes at the cost of additional pre-processing steps and potential artefacts but may be beneficial where different pre-processing methods or parameters are to be evaluated.
 v. Segmentation of microstructures with heterogeneity from the sub-µm to mm-scale scales enables easy area fraction measurements across large panoramic images. Our results also highlight the

vi. uncertainty of such measurements and the importance of choosing and stating the considered image size and placement carefully.
vi. The segmented and therefore discretized microstructure also provides data that can be directly used as input for modelling at the microstructural scale and allows the extraction of detailed geometric data about many features, such as deformation-induced damage sites and how these relate to the surrounding phase distribution.

# 6  Acknowledgment

The authors gratefully acknowledge funding by the Deutsche Forschungsgemeinschaft through projects B2 and T02 of the Collaborative Research Center TRR 188 (project ID 278868966). This work is also funded by the Federal Ministry of Education and Research (BMBF) and the state of North Rhine-Westphalia as part of the NHR Program. Calculations were performed with computing resources granted by RWTH Aachen University under project rwth0535. We further thank Professor Dr. H. Springer for providing the heat-treated samples as well as Mr. M.A. Wollenweber for providing the SEM image of the cross-sectioned sample.

# 7  Code and data availability:

The microstructural image segmentation code, including the trained networks, and the data, including the image masks, for the use by other researchers is in a private Zenodo repository of the authors and can be made publicly available upon paper publication or on prior request.

# 8 Bibliography

[1] F.N. Rhines, Microstructure-property relationships in materials, Metallurgical Transactions A 8 (1977) 127-133.
[2] IEA, Iron and Steel Technology Roadmap—Towards More Sustainable Steelmaking, (2020).
[3] M. Benakis, D. Costanzo, A. Patran, Current mode effects on weld bead geometry and heat affected zone in pulsed wire arc additive manufacturing of Ti-6-4 and Inconel 718, Journal of Manufacturing Processes 60 (2020) 61-74.
[4] H. Xu, Y. Li, C. Brinson, W. Chen, A Descriptor-Based Design Methodology for Developing Heterogeneous Microstructural Materials System, Journal of Mechanical Design 136(5) (2014).
[5] T.M. Nunes, V.H.C. De Albuquerque, J.P. Papa, C.C. Silva, P.G. Normando, E.P. Moura, J.M.R. Tavares, Automatic microstructural characterization and classification using artificial intelligence techniques on ultrasound signals, Expert Systems with Applications 40(8) (2013) 3096-3105.
[6] U. Adiga, M. Gorantla, J. Scott, D. Banks, Y.-S. Choi, Building 3d microstructure database using an advanced metallographic serial sectioning technique and robust 3d segmentation tools, Proceedings of the 2nd World Congress on Integrated Computational Materials Engineering (ICME), Springer, 2013, pp. 243-248.
[7] J. Dargay, D. Gately, M. Sommer, Vehicle ownership and income growth, worldwide: 1960-2030, The energy journal 28(4) (2007).
[8] H. Peregrina-Barreto, I.R. Terol-Villalobos, J.J. Rangel-Magdaleno, A.M. Herrera-Navarro, L.A. Morales-Hernández, F. Manríquez-Guerrero, Automatic grain size determination in microstructures using image processing, Measurement 46(1) (2013) 249-258.
[9] P. Shi, M. Duan, L. Yang, W. Feng, L. Ding, L. Jiang, An Improved U-Net Image Segmentation Method and Its Application for Metallic Grain Size Statistics, Materials (Basel) 15(13) (2022).
[10] X. Li, X. Ma, S. Subramanian, C. Shang, R. Misra, Influence of prior austenite grain size on martensite–austenite constituent and toughness in the heat affected zone of 700 MPa high strength linepipe steel, Materials Science and Engineering: A 616 (2014) 141-147.
[11] J. Tiley, G. Viswanathan, R. Srinivasan, R. Banerjee, D. Dimiduk, H. Fraser, Coarsening kinetics of γ′ precipitates in the commercial nickel base Superalloy René 88 DT, Acta Materialia 57(8) (2009) 2538-2549.
[12] S. Liu, Y.C. Shin, Additive manufacturing of Ti6Al4V alloy: A review, Materials & Design 164 (2019) 107552.
[13] J.P. MacSleyne, J.P. Simmons, M. De Graef, On the use of 2-D moment invariants for the automated classification of particle shapes, Acta Materialia 56(3) (2008) 427-437.
[14] J. Villavicencio, N. Ulloa, L. Lozada, M. Moreno, L. Castro, The role of non-metallic Al2O3 inclusions, heat treatments and microstructure on the corrosion resistance of an API 5L X42 steel, Journal of Materials Research and Technology 9(3) (2020) 5894-5911.
[15] I. Konovalenko, P. Maruschak, O. Prentkovskis, R. Junevičius, Investigation of the rupture surface of the titanium alloy using convolutional neural networks, Materials 11(12) (2018) 2467.
[16] S. Tsopanidis, R.H. Moreno, S. Osovski, Toward quantitative fractography using convolutional neural networks, Engineering Fracture Mechanics 231 (2020) 106992.
[17] T. Falk, D. Mai, R. Bensch, Ö. Çiçek, A. Abdulkadir, Y. Marrakchi, A. Böhm, J. Deubner, Z. Jäckel, K. Seiwald, U-Net: deep learning for cell counting, detection, and morphometry, Nature methods 16(1) (2019) 67-70.
[18] T.C. Hollon, B. Pandian, A.R. Adapa, E. Urias, A.V. Save, S.S.S. Khalsa, D.G. Eichberg, R.S. D'Amico, Z.U. Farooq, S. Lewis, Near real-time intraoperative brain tumor diagnosis using stimulated Raman histology and deep neural networks, Nature medicine 26(1) (2020) 52-58.
[19] M. Cheriet, J.N. Said, C.Y. Suen, A recursive thresholding technique for image segmentation, IEEE transactions on image processing 7(6) (1998) 918-921.

[20] K.V. Mardia, T. Hainsworth, A spatial thresholding method for image segmentation, IEEE transactions on pattern analysis and machine intelligence 10(6) (1988) 919-927.
[21] D. Kim, S. Lee, W. Hong, H. Lee, S. Jeon, S. Han, J. Nam, Image segmentation for FIB-sem serial sectioning of a Si/C–graphite composite anode microstructure based on preprocessing and global thresholding, Microscopy and Microanalysis 25(5) (2019) 1139-1154.
[22] V.S.S.A. Karra, A.K. Verma, A. Guzel, A. Huck, A.D. Rollett, Quantification of alpha lath in Ti-6Al-4V using OpenCV, Materials Characterization 186 (2022) 111802.
[23] F.Y. Shih, S. Cheng, Automatic seeded region growing for color image segmentation, Image and vision computing 23(10) (2005) 877-886.
[24] R. Adams, L. Bischof, Seeded region growing, IEEE Transactions on pattern analysis and machine intelligence 16(6) (1994) 641-647.
[25] S. Beucher, F. Meyer, The morphological approach to segmentation: the watershed transformation, Mathematical morphology in image processing, CRC Press2018, pp. 433-481.
[26] S.S. Al-Amri, N. Kalyankar, S. Khamitkar, Image segmentation by using edge detection, International journal on computer science and engineering 2(3) (2010) 804-807.
[27] S. Savant, A review on edge detection techniques for image segmentation, International Journal of Computer Science and Information Technologies 5(4) (2014) 5898-5900.
[28] G.B. Coleman, H.C. Andrews, Image segmentation by clustering, Proceedings of the IEEE 67(5) (1979) 773-785.
[29] S.M. Azimi, D. Britz, M. Engstler, M. Fritz, F. Mücklich, Advanced steel microstructural classification by deep learning methods, Scientific reports 8(1) (2018) 2128.
[30] C. Kusche, T. Reclik, M. Freund, T. Al-Samman, U. Kerzel, S. Korte-Kerzel, Large-area, high-resolution characterisation and classification of damage mechanisms in dual-phase steel using deep learning, PloS one 14(5) (2019) e0216493.
[31] T. Strohmann, K. Bugelnig, E. Breitbarth, F. Wilde, T. Steffens, H. Germann, G. Requena, Semantic segmentation of synchrotron tomography of multiphase Al-Si alloys using a convolutional neural network with a pixel-wise weighted loss function, Scientific reports 9(1) (2019) 19611.
[32] R. Lorenzoni, I. Curosu, S. Paciornik, V. Mechtcherine, M. Oppermann, F. Silva, Semantic segmentation of the micro-structure of strain-hardening cement-based composites (SHCC) by applying deep learning on micro-computed tomography scans, Cement and Concrete Composites 108 (2020) 103551.
[33] E.A. Holm, R. Cohn, N. Gao, A.R. Kitahara, T.P. Matson, B. Lei, S.R. Yarasi, Overview: Computer Vision and Machine Learning for Microstructural Characterization and Analysis, Metallurgical and Materials Transactions A 51(12) (2020) 5985-5999.
[34] T. Stan, Z.T. Thompson, P.W. Voorhees, Optimizing convolutional neural networks to perform semantic segmentation on large materials imaging datasets: X-ray tomography and serial sectioning, Materials Characterization 160 (2020) 110119.
[35] K. Bugelnig, H. Germann, T. Steffens, B. Plank, F. Wilde, E. Boller, G. Requena, Optimized Segmentation of the 3D Microstructure in Cast Al-Si Piston Alloys, Practical Metallography 55(4) (2018) 223-243.
[36] F. Ajioka, Z.-L. Wang, T. Ogawa, Y. Adachi, Development of High Accuracy Segmentation Model for Microstructure of Steel by Deep Learning, ISIJ International 60(5) (2020) 954-959.
[37] S. Medghalchi, C.F. Kusche, E. Karimi, U. Kerzel, S. Korte-Kerzel, Damage analysis in dual-phase steel using deep learning: transfer from uniaxial to biaxial straining conditions by image data augmentation, Jom 72 (2020) 4420-4430.
[38] S. Medghalchi, E. Karimi, S.-H. Lee, B. Berkels, U. Kerzel, S. Korte-Kerzel, Three-Dimensional Characterisation of Deformation-induced Damage in Dual Phase Steel using Deep Learning, Materials & Design (2023) 112108.
[39] A.R. Durmaz, M. Muller, B. Lei, A. Thomas, D. Britz, E.A. Holm, C. Eberl, F. Mucklich, P. Gumbsch, A deep learning approach for complex microstructure inference, Nat Commun 12(1) (2021) 6272.

[40] D. Iren, M. Ackermann, J. Gorfer, G. Pujar, S. Wesselmecking, U. Krupp, S. Bromuri, Aachen-Heerlen annotated steel microstructure dataset, Sci Data 8(1) (2021) 140.
[41] B.L. DeCost, B. Lei, T. Francis, E.A. Holm, High Throughput Quantitative Metallography for Complex Microstructures Using Deep Learning: A Case Study in Ultrahigh Carbon Steel, Microsc Microanal 25(1) (2019) 21-29.
[42] A. Thomas, A.R. Durmaz, T. Straub, C. Eberl, Automated quantitative analyses of fatigue-induced surface damage by deep learning, Materials 13(15) (2020) 3298.
[43] J. Luengo, R. Moreno, I. Sevillano, D. Charte, A. Peláez-Vegas, M. Fernández-Moreno, P. Mesejo, F. Herrera, A tutorial on the segmentation of metallographic images: Taxonomy, new MetalDAM dataset, deep learning-based ensemble model, experimental analysis and challenges, Information Fusion 78 (2022) 232-253.
[44] C. Shen, X. Wei, C. Wang, W. Xu, A deep learning method for extensible microstructural quantification of DP steel enhanced by physical metallurgy-guided data augmentation, Materials Characterization 180 (2021) 111392.
[45] T.M. Ostormujof, R.P.R. Purohit, S. Breumier, N. Gey, M. Salib, L. Germain, Deep Learning for automated phase segmentation in EBSD maps. A case study in Dual Phase steel microstructures, Materials Characterization 184 (2022) 111638.
[46] L. Fu, H. Yu, M. Shah, J. Simmons, S. Wang, Crystallographic Symmetry for Data Augmentation in Detecting Dendrite Cores, Electronic Imaging 32(14) (2020) 248-1-248-7.
[47] C. Wang, D. Shi, S. Li, A study on establishing a microstructure-related hardness model with precipitate segmentation using deep learning method, Materials 13(5) (2020) 1256.
[48] N.M. Senanayake, J.L. Carter, Computer vision approaches for segmentation of nanoscale precipitates in nickel-based superalloy IN718, Integrating Materials and Manufacturing Innovation 9 (2020) 446-458.
[49] J. Stuckner, B. Harder, T.M. Smith, Microstructure segmentation with deep learning encoders pre-trained on a large microscopy dataset, npj Computational Materials 8(1) (2022).
[50] N. Lutsey, Review of technical literature and trends related to automobile mass-reduction technology, (2010).
[51] C.C. Tasan, M. Diehl, D. Yan, M. Bechtold, F. Roters, L. Schemmann, C. Zheng, N. Peranio, D. Ponge, M. Koyama, An overview of dual-phase steels: advances in microstructure-oriented processing and micromechanically guided design, Annual Review of Materials Research 45 (2015) 391-431.
[52] N. Peixinho, N. Jones, A. Pinho, Application of Dual-Phase and TRIP Steels on the Improvement of Crashworthy Structures, Materials Science Forum 502 (2005) 181-188.
[53] A. Mayyas, A. Qattawi, M. Omar, D. Shan, Design for sustainability in automotive industry: A comprehensive review, Renewable and sustainable energy reviews 16(4) (2012) 1845-1862.
[54] S. Medghalchi, M. Zubair, E. Karimi, S. Sandlöbes-Haut, U. Kerzel, S. Korte-Kerzel, Determination of the Rate Dependence of Damage Formation in Metallic-Intermetallic Mg–Al–Ca Composites at Elevated Temperature using Panoramic Image Analysis, Advanced Engineering Materials 25(21) (2023) 2300956.
[55] S.Y. Allain, I. Pushkareva, J. Teixeira, M. Gouné, C. Scott, Dual-phase steels: the first family of advanced high strength steels, Elsevier, 2020.
[56] A. Garcia-Garcia, S. Orts-Escolano, S. Oprea, V. Villena-Martinez, P. Martinez-Gonzalez, J. Garcia-Rodriguez, A survey on deep learning techniques for image and video semantic segmentation, Applied Soft Computing 70 (2018) 41-65.
[57] A.M. Hafiz, G.M. Bhat, A survey on instance segmentation: state of the art, International journal of multimedia information retrieval 9(3) (2020) 171-189.
[58] S. Minaee, Y. Boykov, F. Porikli, A. Plaza, N. Kehtarnavaz, D. Terzopoulos, Image Segmentation Using Deep Learning: A Survey, IEEE Trans Pattern Anal Mach Intell 44(7) (2022) 3523-3542.
[59] T.M. Inc., MATLAB, Image Labeler application, The MathWorks Inc., Natick, Massachusetts, United States, 2018.
[60] O. Ronneberger, P. Fischer, T. Brox, U-net: Convolutional networks for biomedical image segmentation, Medical Image Computing and Computer-Assisted Intervention–MICCAI 2015: 18th

International Conference, Munich, Germany, October 5-9, 2015, Proceedings, Part III 18, Springer, 2015, pp. 234-241.
[61] M. Drozdzal, E. Vorontsov, G. Chartrand, S. Kadoury, C. Pal, The importance of skip connections in biomedical image segmentation, International Workshop on Deep Learning in Medical Image Analysis, International Workshop on Large-Scale Annotation of Biomedical Data and Expert Label Synthesis, Springer, 2016, pp. 179-187.
[62] F. Isensee, From Manual to Automated Design of Biomedical Semantic Segmentation Methods, 2020.
[63] F. Sultana, A. Sufian, P. Dutta, Evolution of image segmentation using deep convolutional neural network: A survey, Knowledge-Based Systems 201 (2020) 106062.
[64] P. Bilic, P. Christ, H.B. Li, E. Vorontsov, A. Ben-Cohen, G. Kaissis, A. Szeskin, C. Jacobs, G.E.H. Mamani, G. Chartrand, The liver tumor segmentation benchmark (lits), Medical Image Analysis 84 (2023) 102680.
[65] N. Heller, F. Isensee, K.H. Maier-Hein, X. Hou, C. Xie, F. Li, Y. Nan, G. Mu, Z. Lin, M. Han, The state of the art in kidney and kidney tumor segmentation in contrast-enhanced CT imaging: Results of the KiTS19 challenge, Medical image analysis 67 (2021) 101821.
[66] F. Isensee, P.F. Jaeger, S.A.A. Kohl, J. Petersen, K.H. Maier-Hein, nnU-Net: a self-configuring method for deep learning-based biomedical image segmentation, Nat Methods 18(2) (2021) 203-211.
[67] A.C. Wilson, R. Roelofs, M. Stern, N. Srebro, B. Recht, The marginal value of adaptive gradient methods in machine learning, Advances in neural information processing systems 30 (2017).
[68] Q. Wang, Y. Ma, K. Zhao, Y. Tian, A comprehensive survey of loss functions in machine learning, Annals of Data Science (2020) 1-26.
[69] J. Ma, J. Chen, M. Ng, R. Huang, Y. Li, C. Li, X. Yang, A.L. Martel, Loss odyssey in medical image segmentation, Medical Image Analysis 71 (2021) 102035.
[70] I. Goodfellow, J. Pouget-Abadie, M. Mirza, B. Xu, D. Warde-Farley, S. Ozair, A. Courville, Y. Bengio, Generative adversarial networks, Communications of the ACM 63(11) (2020) 139-144.
[71] M. Mirza, S. Osindero, Conditional generative adversarial nets, arXiv preprint arXiv:1411.1784 (2014).
[72] P. Isola, J.-Y. Zhu, T. Zhou, A.A. Efros, Image-to-image translation with conditional adversarial networks, Proceedings of the IEEE conference on computer vision and pattern recognition, 2017, pp. 1125-1134.
[73] M. Ester, H.-P. Kriegel, J. Sander, X. Xu, A density-based algorithm for discovering clusters in large spatial databases with noise, kdd, 1996, pp. 226-231.
[74] J.-C. Yen, F.-J. Chang, S. Chang, A new criterion for automatic multilevel thresholding, IEEE Transactions on Image Processing 4(3) (1995) 370-378.
[75] B.C. Cameron, C.C. Tasan, Towards physical insights on microstructural damage nucleation from data analytics, Computational Materials Science 202 (2022) 110627.
[76] F. Bridier, P. Villechaise, J. Mendez, Analysis of the different slip systems activated by tension in a α/β titanium alloy in relation with local crystallographic orientation, Acta Materialia 53(3) (2005) 555-567.
[77] K.S. Kumar, S. Suresh, M.F. Chisholm, J.A. Horton, P. Wang, Deformation of electrodeposited nanocrystalline nickel, Acta Materialia 51(2) (2003) 387-405.
[78] A. Tkach, N. Fonshtein, V. Simin'kovich, A. Bortsov, Y.N. Lenets, Fatigue crack growth in a dual-phase ferritic-martensitic steel, Soviet materials science: a transl. of Fiziko-khimicheskaya mekhanika materialov/Academy of Sciences of the Ukrainian SSR 20 (1985) 448-453.
[79] A. Bag, K. Ray, E. Dwarakadasa, Influence of martensite content and morphology on the toughness and fatigue behavior of high-martensite dual-phase steels, Metallurgical and Materials Transactions A 32 (2001) 2207-2217.
[80] H. Ghadbeigi, C. Pinna, S. Celotto, J. Yates, Local plastic strain evolution in a high strength dual-phase steel, Materials Science and Engineering: A 527(18-19) (2010) 5026-5032.
[81] K.K. Alaneme, Fracture Toughness (K1C) evaluation for dual phase medium carbon low alloy steels using circumferential notched tensile (CNT) specimens, Materials research 14 (2011) 155-160.

[82] M. Calcagnotto, Y. Adachi, D. Ponge, D. Raabe, Deformation and fracture mechanisms in fine-and ultrafine-grained ferrite/martensite dual-phase steels and the effect of aging, Acta Materialia 59(2) (2011) 658-670.
[83] L. Godefroid, M. Andrade, F. Machado, W. Horta, Effect of prestrain and bake hardening heat treatment on fracture toughness and fatigue crack growth resistance of two dual-phase steels, Proceedings of the Materials Science and Technology 2011 Conference, AIST/ASM, 2011.
[84] M. Guan, H. Yu, Fatigue crack growth behaviors in hot-rolled low carbon steels: A comparison between ferrite–pearlite and ferrite–bainite microstructures, Materials Science and Engineering: A 559 (2013) 875-881.
[85] S. Li, Y. Kang, S. Kuang, Effects of microstructure on fatigue crack growth behavior in cold-rolled dual phase steels, Materials Science and Engineering: A 612 (2014) 153-161.
[86] A.-P. Pierman, O. Bouaziz, T. Pardoen, P. Jacques, L. Brassart, The influence of microstructure and composition on the plastic behaviour of dual-phase steels, Acta Materialia 73 (2014) 298-311.
[87] S. Sun, M. Pugh, Properties of thermomechanically processed dual-phase steels containing fibrous martensite, Materials Science and Engineering: A 335(1-2) (2002) 298-308.
[88] M. Azuma, S. Goutianos, N. Hansen, G. Winther, X. Huang, Effect of hardness of martensite and ferrite on void formation in dual phase steel, Materials Science and Technology 28(9-10) (2012) 1092-1100.
[89] C. Tian, C. Kirchlechner, The fracture toughness of martensite islands in dual-phase DP800 steel, Journal of Materials Research 36 (2021) 2495-2504.
[90] Q. Lai, O. Bouaziz, M. Gouné, L. Brassart, M. Verdier, G. Parry, A. Perlade, Y. Bréchet, T. Pardoen, Damage and fracture of dual-phase steels: Influence of martensite volume fraction, Materials Science and Engineering: A 646 (2015) 322-331.
[91] D.L. Steinbrunner, D. Matlock, G. Krauss, Void formation during tensile testing of dual phase steels, Metallurgical transactions A 19 (1988) 579-589.
[92] G. Lúcio de Faria, L.B. Godefroid, I.P. Nunes, J. Carlos de Lacerda, Effect of martensite volume fraction on the mechanical behavior of an UNS S41003 dual-phase stainless steel, Materials Science and Engineering: A 797 (2020).
[93] H. Shen, T. Lei, J. Liu, Microscopic deformation behaviour of martensitic–ferritic dual-phase steels, Materials science and technology 2(1) (1986) 28-33.
[94] M. Ahmadi, M. Sadighi, H.H. Toudeshky, Computational microstructural model of ordinary state-based Peridynamic theory for damage mechanisms, void nucleation, and propagation in DP600 steel, Engineering Fracture Mechanics 247 (2021) 107660.
[95] K. Ismail, A. Perlade, P.J. Jacques, T. Pardoen, L. Brassart, Impact of second phase morphology and orientation on the plastic behavior of dual-phase steels, International Journal of Plasticity 118 (2019) 130-146.
[96] F.M. Al-Abbasi, J.A. Nemes, Micromechanical modeling of the effect of particle size difference in dual phase steels, International Journal of Solids and Structures 40(13-14) (2003) 3379-3391.
[97] G. A.-F., H. Flower, T. Lindley, Electron backscattering diffraction study of acicular ferrite, bainite, and martensite steel microstructures, Materials Science and Technology 16(1) (2000) 26-40.
[98] L. Ryde, Application of EBSD to analysis of microstructures in commercial steels, Materials Science and Technology 22(11) (2006) 1297-1306.
[99] J.Y. Kang, S.J. Park, M.B. Moon, Phase analysis on dual-phase steel using band slope of electron backscatter diffraction pattern, Microsc Microanal 19 Suppl 5 (2013) 13-6.
[100] F. Zhang, A. Ruimi, D.P. Field, Phase identification of dual-phase (DP980) steels by electron backscatter diffraction and nanoindentation techniques, Microscopy and microanalysis 22(1) (2016) 99-107.
[101] C. Tian, On the damage initiation in dual phase steels: Quantitative insights from in situ micromechanics, Ruhr-Universität Bochum, 2021.
[102] P.T. Pinard, A. Schwedt, A. Ramazani, U. Prahl, S. Richter, Characterization of dual-phase steel microstructure by combined submicrometer EBSD and EPMA carbon measurements, Microsc Microanal 19(4) (2013) 996-1006.

[103] J.-Y. Kang, H.K. Do, S.-I. Baik, T.-H. Ahn, Y.-W. Kim, H.N. Han, K.H. Oh, H.-C. Lee, S.H. Han, Phase analysis of steels by grain-averaged EBSD functions, ISIJ international 51(1) (2011) 130-136.
[104] C. Tian, D. Ponge, L. Christiansen, C. Kirchlechner, On the mechanical heterogeneity in dual phase steel grades: Activation of slip systems and deformation of martensite in DP800, Acta Materialia 183 (2020) 274-284.
[105] B. Lin, S. Medghalchi, S. Korte-Kerzel, B.-X. Xu, A Machine Learning Enabled Image-data-driven End-to-end Mechanical Field Predictor For Dual-Phase Steel, Pamm 22(1) (2023).
[106] C.F. Kusche, F. Pütz, S. Münstermann, T. Al-Samman, S. Korte-Kerzel, On the effect of strain and triaxiality on void evolution in a heterogeneous microstructure–A statistical and single void study of damage in DP800 steel, Materials Science and Engineering: A 799 (2021) 140332.
[107] C. Du, R. Petrov, M.G.D. Geers, J.P.M. Hoefnagels, Lath martensite plasticity enabled by apparent sliding of substructure boundaries, Materials & Design 172 (2019).
[108] T. Vermeij, C.J.A. Mornout, V. Rezazadeh, J.P.M. Hoefnagels, Martensite plasticity and damage competition in dual-phase steel: A micromechanical experimental–numerical study, Acta Materialia 254 (2023).
[109] L. Morsdorf, O. Jeannin, D. Barbier, M. Mitsuhara, D. Raabe, C.C. Tasan, Multiple mechanisms of lath martensite plasticity, Acta Materialia 121 (2016) 202-214.

# Supplementary Materials

## SM1 - Extraction of angular martensite phase fraction around the damage site

In order to extract the angular profile of each damage site, we first isolate the damage site under consideration using the segmented microstructure (Step 1 in Figure SM 1). Then, in Step 2, we expand the footprint of each damage by five pixels (i.e., 16 nm) outwards from the edge of the damage site. Applying an exclusive or (XOR) to this extended region removes the damage site from the area and leaves us with an irregular shaped torus following the shape of the damage site in the segmented micrograph (Step 3).

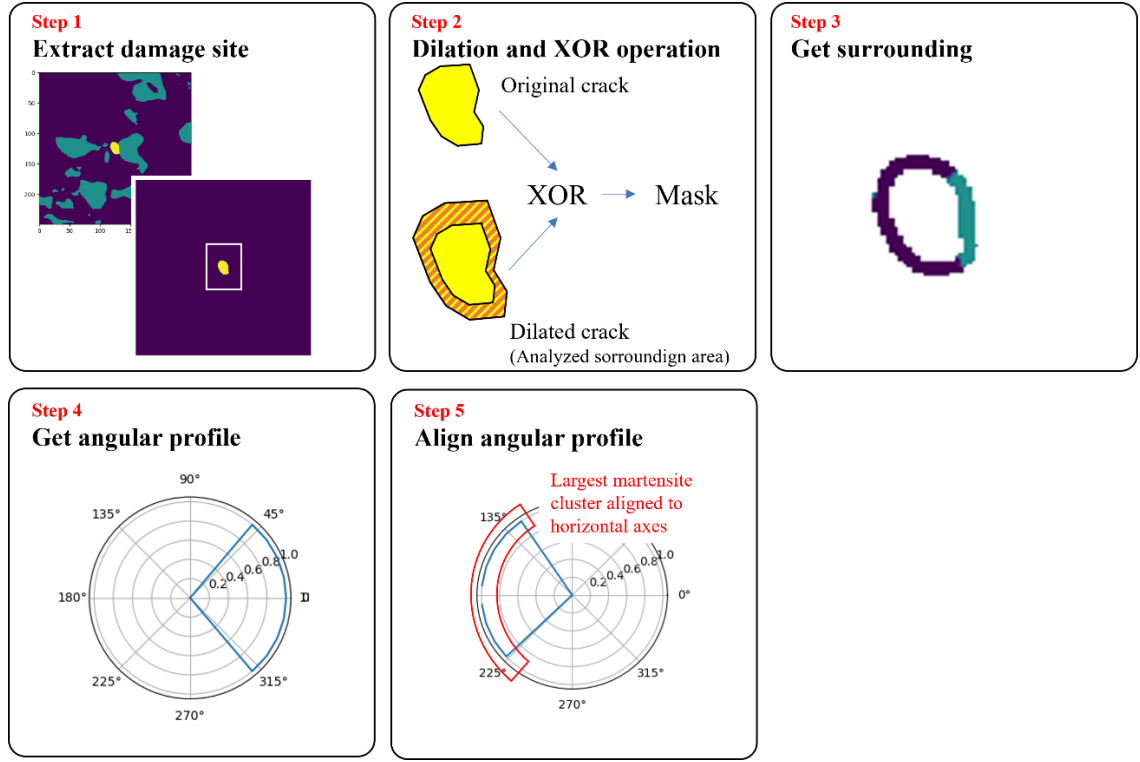

*Figure SM 1. Phase fraction calculation in a donut/Simit shape area around the damage site*

The angular profile (Step 4) is then extracted in the following way: First, the centre of the damage site is determined. This then serves as the origin of a polar coordinate system. As we move around the plane, from zero to 360 degrees, we record the martensite fraction that can be identified from the segmented microstructure. Then, martensite fraction is binarized with the critical value (v=0.15). This ensures sensitive martensite detection. A largest martensite cluster is identified from the procedure described in section SM3. Since a given damage site may be surrounded by multiple martensite islands, we orient the damage such that the largest martensite island is always aligned with the horizontal axis (Step 5) to obtain consistent results. Additionally, the martensite phase fraction $F_m$ in the area surrounding the damage site was measured as $F_m = P_m/P_{tot}$, where $P_m$ is the number of pixels attributed to martensite by the segmentation algorithm, and $P_{tot}$ is the total number of pixels in the selected area around the damage site.

## SM2 - Extraction of the width profile around the damage site

For an easier local martensite islands/channels width calculation, around the damage sites, we rotate the images to ensure that the major axis of the damage site is vertical. To localize the martensite islands, we take the window size of 20 µm × 20 µm around the centroid of the damage sites. Subsequently, we extracted the connected martensite island using iterative binary dilation with modified footprint. A modified footprint allowed the exclusion of irrelevant part of the martensite, as depicted in Figure SM 2. Note that, in Figure SM 2.a unwanted part is highlighted by white box. It is effectively removed in Figure SM 2.f.

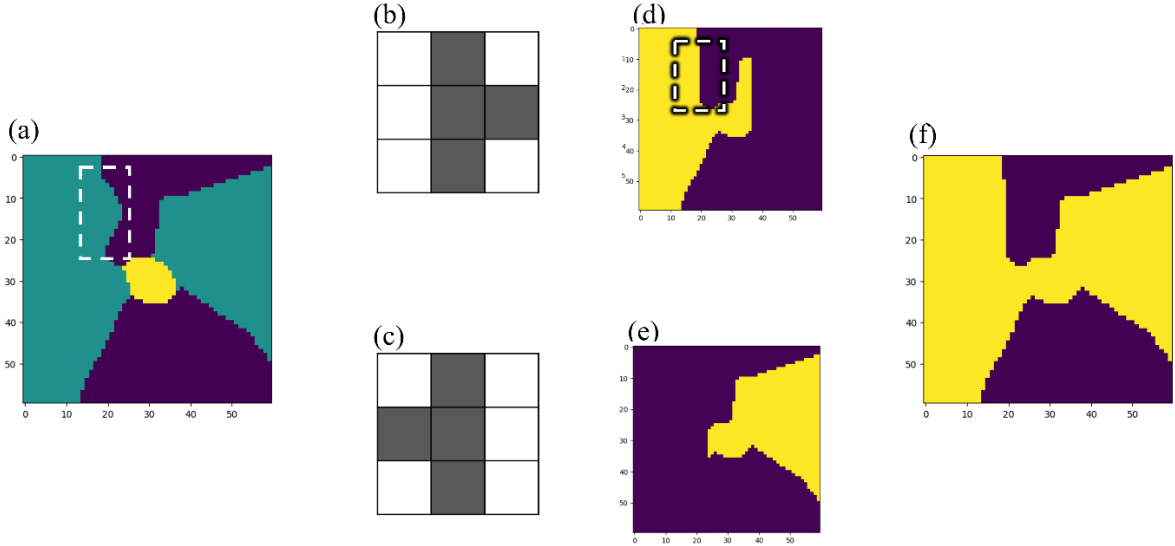

*Figure SM 2. (a) microstructure before dilation, (b, c) binary dilation footprints for left- and right-side dilation, (d, e) result of left- and right-side dilation and (e) final dilation result, by computing OR operation between (d) and (e).*

To achieve the exclusion of the unwanted part of the martensite, first, the crack is dilated with one of the modified footprints shown in Figure SM 2.b or c. Each different footprint is required for the left-side dilation and right-side dilation. Then, an OR operation between dilated crack and martensite cluster is computed. This operation is iterated for ($\sqrt{2}$ × image width) steps to ensure a complete dilation of the crack is achieved (Figure SM 2.d and e). Then, the calculated left-side dilation and right-side dilation are combined by OR operation. Then, the martensite island width profile is obtained by counting the number of martensite pixels. To preserve the asymmetric nature of interface decohesion, martensite width profile flipped horizontally, if necessary, to ensure that the left side always contains more martensite. Figure SM 3, shows an example of this process.

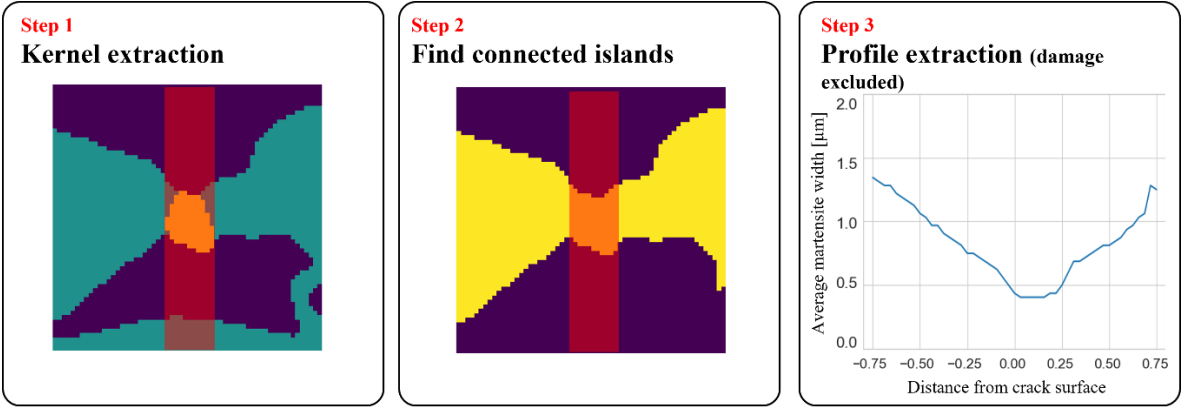

*Figure SM 3.Calculation of the width of surrounding martensite island.*

### SM3 - Number of surrounding martensite islands

The number of martensite clusters was determined by analyzing the martensite angular profile. To ensure sensitivity, the profile was filtered by a thresholder using a critical value of 0.15. This resulted in a binary profile, which was then analyzed to identify consecutive regions of martensite and ferrite, considering periodicity. For example, an array of [1, 0, 1, 1, 0, 1] would yield two martensite islands with sizes of 2 and two ferrite islands with sizes of 1. Then, the number of martensite clusters ($M_C$) was determined by counting the number of unique martensite islands, and the average martensite angular span was calculated using Equation 1.

$$M_{AS} = \frac{M_M}{M_C} \cdot \frac{360°}{M_M + M_F} \qquad \text{Equation 1}$$

Where $M_{AS}$ is martensite average angular span, $M_M$ is the number of martensite pixels in the profile, $M_F$ is the number of ferrite pixels in the profile, and $M_C$ is the martensite island number.